\documentclass[%
 reprint,
 amsmath,amssymb,
 aps,
]{revtex4-2}

\usepackage{graphicx}
\usepackage{dcolumn}
\usepackage{algorithm}
\usepackage{algpseudocode}
\usepackage{bm}

\newtheorem{corollary}{Corollary}
\newtheorem{lemma}{Lemma}
\newtheorem{theorem}{Theorem}

\newtheorem{definition}{Definition}
\newcommand{\Beginproof}{{\em Proof.}  }
\newcommand{\Endproof}{\hfill$\Box$\\}
\begin{document}

\preprint{APS/123-QED}

\title{Implementation of Quantum Fourier Transform and Quantum Hashing for a Quantum Device with Arbitrary Qubits Connection Graphs}

\author{Kamil Khadiev}
\email{kamilhadi@gmail.com}
 \altaffiliation{Institute of Computational Mathematics and Information Technologies, Kazan Federal University, Kazan, Russia}
\author{Aliya Khadieva}%
 \email{aliya.khadi@gmail.com}
\affiliation{%
 Institute of Computational Mathematics and Information Technologies, Kazan Federal University, Kazan, Russia
}%


\author{Zeyu Chen}
\affiliation{%
 School of Mathematical Sciences, Zhejiang University, Hangzhou 310058, People's Republic of China
}%
\author{Junde Wu}
\email{wjd@zju.edu.cn}
\affiliation{
 School of Mathematical Sciences, Zhejiang University, Hangzhou 310058, People's Republic of China
}%


\date{\today}

\begin{abstract}
In the paper, we consider quantum circuits for Quantum fingerprinting (quantum hashing) and quantum Fourier transform (QFT) algorithms. Quantum fingerprinting (quantum hashing) is a well-known technique for comparing large objects using small images. The QFT algorithm is a very popular technique used in many algorithms.
We present a generic method for constructing quantum circuits for these algorithms for quantum devices with restrictions.  Many quantum devices (for example, based on superconductors) have restrictions on applying two-qubit gates. The restrictions are presented by a qubits connection graph.  Typically, researchers consider only the linear nearest neighbor (LNN) architecture, but current devices have more complex graphs. We present a method for arbitrary connected graphs that minimizes the number of CNOT gates in the circuit.

The heuristic version of the method is fast enough and works with $O(n^5)$ time complexity, where $n$ is the number of qubits. The certain version of the algorithm has an exponential time complexity that is $O(n^22^n)$. We compare quantum circuits built by our algorithm with quantum circuits optimized for specific graphs that are Linear-nearest-neighbor (LNN) architecture, ``sun'' (a cycle with tails, presented by 16-qubit IBMQ device) and ``two joint suns'' (two joint cycles with tails, presented by 27-qubit IBMQ device). Our generic method gives similar results with little bit more CNOT gates. At the same time, our method allows us to construct a circuit for arbitrary connected graphs.
\end{abstract}

\maketitle


\section{Introduction}
Quantum Fourier transform (QFT) \cite{k1995} is a well-known computational technique used in many quantum algorithms.
The technique is used in quantum addition \cite{d2000}, quantum phase estimation (QPE) \cite{k1995}, quantum amplitude estimation (QAE)\cite{bhmt2002}, the algorithm for solving linear systems of equations \cite{hhl2009}, Shor’s factoring algorithm \cite{s1999}, and others. 

In this paper, we are interested in the implementation of this technique on quantum devices as a quantum circuit with basic gates. We are focusing on minimization of two-qubit quantum gates in such a circuit because they are the most ``expensive'' to implement. Many types quantum computers (for example, quantum devices based on superconductors) do not allow us to apply two-qubit gates for arbitrary pair of qubits, but have a graph that represents such a restriction. Vertices of the graph correspond to qubits, and two-qubit gates can be applied only to qubits corresponding to vertices connected by an edge. We say that such a graph represents qubits topology for a device. In this paper, we focus on the number of CNOT gates in a quantum circuit for the QFT algorithm with respect to such a graph. Let a CNOT cost of a circuit be the number of CNOT gates in the circuit. The CNOT cost for a linear nearest-neighbor (LNN) architecture (where the graph is just a chain) was explored by Park and Ahn in \cite{park2023reducing}. They presented a circuit that has $n^2+n-4$ CNOT cost, where $n$ is the number of qubits. It improved previous results of \cite{nc2010,fdh2004,swd2011,kds2017,bbwdr2019,best1996,tko2007,pa2022}. At the same time, as the authors mentioned, their technique cannot be generalized for more complex graphs. Khadieva in \cite{k2024aliya} suggested a quantum circuit for more complex architecture that is a cycle with tails (like a ``sun'' or ``two joint suns''). At the same time, the CNOT cost of this circuit is $1.5n^2$. 

Here, we present a general method that allows us to develop a quantum circuit of the QFT algorithm for an arbitrary connected graph that represents a qubits connection graph. Our solution is based on a solution for the Travelling salesman problem (TSP) for a modification of the original graph. Here we present a method and an algorithm for constructing the circuit with $O(n2^n)$ time complexity, where $n$ is the number of vertices of the original graph, and approximate solution with polynomial time complexity $O(n^5)$ and the differential approximation ratio $\frac{1}{2}$.

The constructed circuit has the CNOT cost from $1.5n^2-1.5n-1$ to $1.5n^3-1.5n^2-2n$ 
depending on the complexity of the graph. If the graph has a Hamiltonian path, then it has the minimal possible CNOT cost at most $1.5n^2-1.5n-1$. The CNOT cost of graphs without this property can be up to $1.5n^3-1.5n^2-2n$. Additionally, we compare our results with circuits for specific graphs. In the case of LNN, the CNOT cost is $1.5n^2-1.5n-1$ that is $1.5$ times larger than the result of \cite{park2023reducing}. The circuits built by our approach are larger by $n$ CNOT gates than the circuit of \cite{k2024aliya} with the CNOT cost $1.5n^2-2.5n-1$. For more complex graphs like 16-qubit and 27-qubit Eagle r3 architectures of IBMQ that is a cycle with tails (like a ``sun'') or its modifications, our generic technique gives a CNOT cost a bit larger than the circuit \cite{k2024aliya} that was especially constructed for these architectures. Our circuits have CNOT cost $342$ for 16-qubit architecture and $1009$ for 27- qubit one; and \cite{k2024aliya} has CNOT cost $324$ and $957$, respectively. The difference is about 5\%.

Another technique discussed in this paper is quantum fingerprinting or quantum hashing. 
This technique is well-known and it allows us to compute a short hash or fingerprint that identifies an original large data with high probability. Nature of the technique and its implementation is close to the QFT algorithm \cite{av2020,k2024aliya} 

The probabilistic (randomized) technique was developed by Frievalds \cite{Fre79}. Then,  Ambainis and Frievalds \cite{af98} developed its quantum counterpart for automata that was improved by Ambainis and Nahimovs in \cite{an2008,an2009}. Later Buhrman et al. in \cite{bcwd2001} provided an explicit definition of the quantum fingerprinting for constructing an efficient quantum communication protocol for equality testing. The technique was applied for branching programs by Ablayev, Gainutdinova, Vasiliev and co-authors \cite{agkmp2005,ag05,av2009,av2011,av2013}. Later, they developed the concept of cryptographic quantum hashing \cite{av2013hash,av2014,aav2014,aa2015,aav2016,aavz2016,v2016binary,vlz2017,aakv2018,aav2018,vvl2019,aav2020}. Then, different versions of hashing functions were applied by Ablayev, Vasiliev, Ziiatdinov, Zinnatullin and other researchers \cite{v2016,av2020,z2016,z2016group,z2023}. The technique was also extended for qudits \cite{av2022}.
This approach was widely used in different areas like stream processing algorithms \cite{l2009,l2006}, query model algorithms \cite{aaksv2022}, online algorithms \cite{kk2019disj,kk2022}, branching programs \cite{kk2017,kkk2022,agky16}, developing quantum devices \cite{v2016model}, automata (discussed earlier, \cite{gy2017,YS10A,gy2018,gy2015}) and others.

Turaykhanov, Akat’ev and co-authors implemented the technique in a photon-based real quantum device \cite{avsak2022,tavak2021}. At the same time, the algorithm was embedded into this device. When we discuss the implementation of the algorithm for ``universal'' quantum devices like IBMQ quantum computers or similar, it is important to minimize the number of quantum gates with respect to the architecture restrictions of a quantum device as we discussed for QFT above. The basic implementation part of quantum hashing circuits is a uniformly controlled rotation operator. The CNOT cost of this operator with respect to qubits connection graph previously was discussed in \cite{mottonen2005decompositions} for linear nearest-neighbor (LNN) architecture and in \cite{zkk2023} for more complex graphs.

K\={a}lis in his master thesis \cite{kalis18}  suggested a shallow circuit for the quantum fingerprinting for 3 qubits. Later the approach was developed in detail and presented in a general way for quantum hashing by Ziatdinov, Khadieva, and  Yakary{\i}lmaz \cite{ziiatdinov2023gaps}. The approach allows us to reduce the number of CNOT gates exponentially in computational experiments. At the same time, the theoretical exponential superiority of the shallow circuit is not shown \cite{ziiatdinov2023gaps}. By the way, the method is very perspective for current and near-future quantum devices that definitely cannot support a huge number of CNOT gates. The  efficient implementation of shallow circuit for an automaton that recognizes $MOD_p=\{a^k: k$ mod $p=0\}$ language based on quantum hashing algorithm was propsed by Khadieva, Salehi, and Yakary{\i}lmaz \cite{ksy2024} for devices based on LNN architecture. Later a similar technique was applied by Khadieva \cite{k2024aliya} for more complex architecture that is a cycle with tails (like a ``sun'' and  ``two joint suns'') represented by 16-qubit and 27-qubit Eagle r3 IBMQ architectures. Vasiliev \cite{v2023} discussed a similar method but for Rz gates instead of Ry gates.

In this paper, we present a general method that allows us to develop a quantum circuit of quantum hashing algorithm for an arbitrary qubits connection graph. It uses the same tools and approaches as the algorithm for constructing the circuit of QFT.
 Note that the shallow circuit for quantum fingerprinting and the circuit for QFT have similar structures as it was discussed by Khadieva \cite{k2024aliya}. That is why we consider these two topics together and use common methods for both algorithms.

The CNOT cost of the constructed circuit is from $3n-2$ to $\frac{3}{2}n^2 +\frac{3}{2}n-2$ depends on the complexity of the graph. If the graph has a Hamiltonian path, then it has the minimal possible CNOT cost. If it does not have such a path, then it can reach $\frac{3}{2}n^2 +\frac{3}{2}n-2$. The presented cost value is for one implementation step of the quantum hashing algorithm. If we present the complexity of $\ell$ steps of the algorithm, then the constructed circuit has the CNOT cost from $(3n-4)\ell+2$ to $(\frac{3}{2}n^2+\frac{3}{2}n-4)\ell+2$, which depends on the complexity of the graph. When we apply our approach for LNN architecture, we obtain a circuit with the same CNOT cost as the circuit especially built for LNN  \cite{ksy2024}. The same situation we have if we consider the case of 16-qubit and 27-qubit Eagle r3 IBMQ architectures \cite{k2024aliya}.

The structure of this paper is the following.
Section \ref{sec:prelims} describes required notations and preliminaries. Graph theory tools are presented in Section \ref{sec:tools}.
Section \ref{sec:hash} provides an algorithm for generating quantum circuit for quantum hashing. The circuit for Quantum Fourier Transform is discussed in Section \ref{sec:qft}. The final Section \ref{sec:concl} concludes the paper and contains some open questions. 

\section{Preliminaries}\label{sec:prelims}
\subsection{Graph Theory}
Consider an undirected unweighted graph $G=(V,E)$, where $V$ is the set of vertices, and $E$ is the set of undirected edges. Let $n=|V|$ be the number of vertices, and $m=|E|$ be the number of edges. 

A non-simple path $P$  is a sequence of vertexes $(v_{i_1},\dots,v_{i_h})$ that are connected by edges, i.e. $(v_{i_j},v_{i_{j+1}})\in E$ for all $j\in\{1,\dots,h-1\}$, here $h-1$ is the length of the path. A path is called a non-simple cycle if $v_{i_1}=v_{i_h}$. Note that a non-simple path and a non-simple cycle can contain duplicates.

Let the length of the path be the number of vertices in the path, $len(P)=h$.

A path $P=(v_{i_1},\dots,v_{i_h})$ is called simple if there are no duplicates among $v_{i_1},\dots,v_{i_h}$.  A simple cycle $C=(v_{i_1},\dots,v_{i_h})$ is a non-simple path with only one duplicate $v_{i_1}=v_{i_h}$.
The distance $dist(v,u)$ is the length of the shortest path between vertices $v$ and $u$. Typically if we say just a ``path'' or a ``cycle'', then we mean a ``simple path'' or a ``simple cycle'', respectively. 

Let $\textsc{Neighbors}(v)$ be a list of neighbors for a vertex $v$, i.e. $\textsc{Neighbors}(v)=(u_{i_1},\dots,u_{i_k})$ such that $(v,u_{i_j})\in E$.

Let us consider an undirected weighted graph $S=(V',E')$, where $V'$ is the set of vertices, and $E'$ is the set of undirected weighted edges. Let $w:V'\times V'\to \mathbb{R}$ be a weight function. We assume that $w(v,u)=\infty$ if $(v,u)\not\in E$.

For a path $P=(v_{i_1},\dots,v_{i_h})$, the length  $w(P)$ is the sum of edge weights in the path, i.e. $w(P)=\sum\limits_j^{h-1}w(v_{i_j},v_{i_{j+1}})$.
\subsection{Quantum circuits}\label{sec:qcirc}
Quantum circuits consist of qubits and gates. A state of a qubit is a column-vector from ${\cal H}^2$ Hilbert space. It can be represented by $a_0|0\rangle+a_1|1\rangle$, where $a_0,a_1$ are complex numbers such that $|a_0|^2+|a_1|^2=1$, and $|0\rangle$ and $|1\rangle$ are unit vectors. Here we use Dirac notation. A state of $n$ qubits is represented by a column-vector from ${\cal H}^{2^n}$ Hilbert space. It can be represented by $\sum_{i=0}^{2^n-1}a_i|i\rangle$, where $a_i$ is a complex number such that $\sum_{i=0}^{2^n-1}|a_i|^2=1$, and $|0\rangle,\dots |2^n-1\rangle$ are unit vectors. Graphically, on a circuit, qubits are presented as lines. 

As basic gates, we consider the following ones:

  $H=\frac{1}{\sqrt{2}}\begin{pmatrix}
1 & 1 \\
1 & -1 
\end{pmatrix}$,     $X=\begin{pmatrix}
0 & 1 \\
1 & 0 
\end{pmatrix}$,

 $R_y(\xi)=\begin{pmatrix}
cos(\xi/2) & -sin(\xi/2) \\
sin(\xi/2) & cos(\xi/2) 
\end{pmatrix}$,

$R_z(\xi)=\begin{pmatrix}
e^{\frac{i\xi}{2}} & 0 \\
0 & e^{-\frac{i\xi}{2}} 
\end{pmatrix}$,
$CNOT=\begin{pmatrix}
1 & 0 & 0 & 0\\
0 & 1 & 0 & 0\\
0 & 0 & 0 & 1\\
0 & 0 & 1 & 0 
\end{pmatrix}$.  

Additionally, we consider five non-basic gates

 $R_k=\begin{pmatrix}
1 & 0 \\
0 & e^{\frac{i\pi}{2^{k-1}}} 
\end{pmatrix}$,
$CR_k=\begin{pmatrix}
1 & 0 & 0 & 0\\
0 & 1 & 0 & 0\\
0 & 0 & 1 & 0\\
0 & 0 & 0 & e^{\frac{i\pi}{2^{k-1}}} 
\end{pmatrix}$,

$CR_z(\xi)=\begin{pmatrix}
1 & 0 & 0 & 0\\
0 & 1 & 0 & 0\\
0 & 0 & e^{\frac{i\xi}{2}}  & 0\\
0 & 0 & 0 & e^{-\frac{i\xi}{2}} 
\end{pmatrix}$, $SWAP=\begin{pmatrix}
1 & 0 & 0 & 0\\
0 & 0 & 1 & 0\\
0 & 1 & 0 & 0\\
0 & 0 & 0 & 1 
\end{pmatrix}$,

$CR_y(\xi)=\begin{pmatrix}
1 & 0 & 0 & 0\\
0 & 1 & 0 & 0\\
0 & 0 & cos(\xi/2) & -sin(\xi/2) \\
0 & 0 & sin(\xi/2) & cos(\xi/2) 
\end{pmatrix}$.

A reader can find more information about quantum circuits in \cite{nc2010,aazksw2019part1,k2022lecturenotes}


\section{Tools}\label{sec:tools}
Let us consider an undirected unweighted connected graph $G=(V,E)$ such that $n=|V|$ is a number of vertices and $m=|E|$ is a number of edges.

In this section, we consider the problem of searching for the shortest non-simple path in the graph $G$ that visits all vertices at least once.
Formally, we want to construct a non-simple path $P=(v_{i_1},\dots,v_{i_k})$ such that 
\begin{itemize}
    \item the path visits all vertices, i.e. $\{v_{i_1},\dots,v_{i_k}\}=V$; and
    \item the length of the path $P$ is the minimal possible.
\end{itemize}
The problem is close to the Hamiltonian path problem \cite{cormen2001}, but here we are allowed to visit a vertex several times. That is why any connected graph has the required path.

To solve this problem we construct a new complete weighted graph $S=(V',E')$ such that 
\begin{itemize}
    \item The set of vertices $V'$ is the same as the set of vertices $V$.
    \item Each pair of vertices $u',v'\in V'$ are connected and the weight $w(u',v')$ is the length of the shortest path between corresponding vertices $u$ and $v$ from $V$, i.e. $w(u',v')=dist(u,v)$.
\end{itemize}
We call it ``supergraph''.
Let a TSP path (or a Travelling salesman problem path) $P'$ be the shortest simple path $P'$ in $S$ that is a solution to the Travelling salesman problem  \cite{cormen2001}. It is a path that visits all vertices from $V'$ exactly once and has the minimal possible length. In a similar way, we can define a TSP cycle $C'$.

For a TSP path $P'=(v_{i_1},\dots,v_{i_n})$, let $\alpha(P')$ be a path in the graph $G$ such that \[\alpha(P')=P_{v_{i_1},v_{i_2}}\circ \tilde{P}_{v_{i_2},v_{i_3}}\circ \tilde{P}_{v_{i_3},v_{i_4}}\circ\dots\circ \tilde{P}_{v_{i_{n-1}},v_{i_n}}.\]
Here $P_{v_{i_1},v_{i_2}}$ is the shortest path in $G$ between $v_{i_1}$ and $v_{i_2}$; the path $\tilde{P}_{v_{i_j},v_{i_{j+1}}}$  is the shortest path in $G$ between $v_{i_j}$ and $v_{i_{j+1}}$ excluding the first vertex in the path that is $v_{i_j}$; $\circ$ is the concatenation operation.

A TSP path for $S$ and the shortest non-simple path in $G$ that visits all vertices at least once has the following connection.

\begin{lemma}\label{lm:tsp}
The shortest non-simple path $P$ in $G$ that visits all vertices can be obtained by a TSP path $P'$ for $S$ as $P=\alpha(P')$.
\end{lemma}
\Beginproof
Let $P=\alpha(P')$. Assume that there is another non-simple path $\hat{P}$ that is shorter than $P$ and visits all vertices of $G$ at least once, i.e. $len(\hat{P})<len(P)$. Let $(i_1,\dots,i_n)$ be the permutation of $(1,\dots,n)$ that presents the order of visiting vertices by $\hat{P}$ for the first time. The length of the simple path $\hat{P}'=(v'_{i_1},\dots,v'_{i_n})$ in $S$ is less or equal to the length of the non-simple path $\hat{P}$ in $G$, i.e. $len(\hat{P})\geq w(\hat{P}')$. It happens because the weight of an edge in $S$ is the length of the shortest path between corresponding vertices of $G$, but $\hat{P}$ visit more vertices (is not the shorts).

At the same time, $\hat{P}'$ visits all vertices of $S$. So, $w(P')\leq w(\hat{P}')$ because $P'$ is a TSP path.

Summarizing, $len(P)=w(P')\leq w(\hat{P}')\leq len(\hat{P})$. It contradicts our assumption $len(P)>len(\hat{P})$.
\Endproof
There is a similar connection for cycles also.
\begin{lemma}
The shortest non-simple cycle $C$ in $G$ that visits all vertices can be obtained by a TSP cycle $C'$ for $S$ as $C=\alpha(C')$.
\end{lemma}
\Beginproof
The proof is the same as for Lemma \ref{lm:tsp}.
\Endproof

We can note the following property:
\begin{lemma}\label{lm:path-len}
The length of the shortest non-simple path $P$ that visits all vertices in $G$ at least once is such that $n\leq len(P)\leq 0.5n^2+0.5n$. A similar result is for a cycle $C$, i.e. $n\leq len(C)\leq 0.5n^2+1.5n-1$
\end{lemma}
\Beginproof
The length of the shortest path in $G$ for any vertices $u$ and $v$ is $1\leq dist(u,v)\leq n$ for $u,v\in V$. At the same time, $w(u',v')=dist(u,v)$ for corresponding $u'$ and $v'$ from $V'$. The length $len(P)=w(P')$ where $P=\alpha(P')$. The length $w(P')$ is the sum of weights of $n-1$ edges. 

Let us look at the path $\alpha(P)$. If we visit a vertex $v_i$ for the first time and after some steps, we visit a vertex $v_j$ for the first time, then in the worst case, between $v_j$ and $v_i$ we have only already visited vertices.

If $j_1,\dots,j_n$ are the indexes of the first time visiting vertices, then the worst case is $j_1=1$, $j_2-j_1=1$, $j_3-j_2=2$,\dots, $j_i-j_{i-1}=i-1$, $j_n-j_{n-1}=n-2$. In that case, $k=1+2+3+\dots+n-1=n(n+1)/2$. 
So, we have $1\cdot (n-1) \leq w(P')\leq n(n+1)/2=0.5n^2+0.5n$.

For a cycle, $w(C')$ is the sum of weights of $n$ edges, and the distance between the last two pairs of vertices can be $n-1$ and the inequality is  $1\cdot n \leq w(C')\leq 0.5n^2+1.5n-1$. 
\Endproof

Based on this connection, we can suggest an algorithm for searching a shortest non-simple path in the graph $G$ that visits all vertices at least once.

Assume that we have a procedure that finds a TSP path for $S$ that is $\textsc{TSPpath}(S)$. As an algorithm, we can use the Bellman-Held-Karp algorithm \cite{b62,hk62}. The algorithm has $O(n^22^n)$ time complexity.

Additionally, we have a procedure $\textsc{ShortestPaths}(G)$ that constructs two $n\times n$-matrices $W$ and $A$ by a graph $G$. 

Elements of the matrix $W$ are lengths of the shortest paths between each pair of vertices in $G$, i.e. $W[v,u]=dist(v,u)$. We can use $W$ as a weight matrix for $S$. In other words, $w(v',u')=W[v,u]$, where $v',u'\in V'$, and $v$ and $u$ are corresponding vertices from $V$.

The matrix $A$ represents the shortest paths between vertices of $G$. The element $A[v,u]$ is the last vertex in the shortest path between $v$ and $u$. In other words, if $t=A[v,u]$, then $P_{v,u}=P_{v,t}\circ u$. Based on this fact, we can present a procedure $\textsc{GetPath}(v,u)$ that computes $P_{v,u}$ using the matrix $A$. The implementation of the procedure is presented in Algorithm \ref{alg:getpath}.

\begin{algorithm}[H]
\caption{Implementation of $\textsc{GetPath}(v,u)$}\label{alg:getpath}
\begin{algorithmic}
\State $t\gets A[v,u]$
\State $P_{v,u}\gets (u)$
\While{$t\neq v$}
\State $P_{v,u}\gets (t)\circ P_{v,u} $
\State $t\gets  A[v,t]$
\EndWhile
\State $P_{v,u}\gets (v)\circ P_{v,u}$
\State \Return $P_{v,u}$
\end{algorithmic}
\end{algorithm}

Let the procedure $\textsc{GetPathNoFirst}(v,u)$ returns $\tilde{P}_{v,u}$ that is the path $P_{v,u}$ without the first element. The implementation is the same, but without the $P_{v,u}\gets (v)\circ P_{v,u}$ line.

We can construct these two matrices using $n$ invocations of the Breadth First Search (BFS) algorithm \cite{cormen2001}. The total time complexity for the matrices construction is $O(n^3)$. The algorithm for constructing $A$ and $W$ is presented in Appendix \ref{apx:floyd} for completeness of presentation.

Due to $V'=V$, and the matrix $W$ represents weights of the supergraph's edges, we can say that $\textsc{ShortestPaths}(G)$ constructs the graph $S$.

Finally, we can present Algorithm \ref{alg:tsp} for solving the problem.

\begin{algorithm}[H]
\caption{Implementation Algorithm for  searching a shortest non-simple path in the graph $G$ that visits all vertices at least once}\label{alg:tsp}
\begin{algorithmic}
\State $W,A\gets \textsc{ShortestPaths}(G)$\Comment{$W$ and $V$ define $S$}
\State $(v_{i_1},\dots,v_{i_k})=P'\gets\textsc{TSPpath}(S)$
\State $\alpha(P')\gets \textsc{GetPath}(v_{i_1},v_{i_2})$
\For{$j\in\{2,\dots,k-1\}$}
\State $\alpha(P')\gets \alpha(P')\circ \textsc{GetPathNoFirst}(v_{i_j},v_{i_{j+1}})$
\EndFor
\end{algorithmic}
\end{algorithm}

Note that if we replace the procedure $\textsc{TSPpath}(S)$ by the procedure  $\textsc{TSPcycle}(S)$ that finds a TSP cycle, then we find the shortest non-simple cycle in $G$ that visits all vertices at least once. The time complexity of $\textsc{TSPcycle}(S)$ is $O(n2^n)$.
We can also modify the  $\textsc{TSPpath}(S)$ such that it finds a path from fixed starting vertex $v_{start}$ that is $\textsc{TSPpath}(S,v_{start})$. In that case, the time complexity is $O(n2^n)$.

\begin{theorem}\label{th:path-compl}
The time complexity of the presented solution for searching a shortest non-simple path that visits all vertices at least once in the graph $G=(V,E)$ is $O(n^22^n)$, where $n=|V|$. In the case of a cycle that visits all vertices and a path with a fixed starting vertex, the time complexity is $O(n2^n)$.  
\end{theorem}

\paragraph{Remark}
We can solve the same problem using quantum algorithms. For $\textsc{TSPpath}(S)$, $\textsc{TSPpath}(S,v_{start})$  and $\textsc{TSPcycle}(S)$, we can use quantum analogue of Bellman-Held-Karp algorithm presented by Ambainis et.al \cite{abikpv2019}. For  $\textsc{ShortestPaths}(S)$, we can use the quantum version of the BFS algorithm \cite{ll2015,ll2016,dhhm2006} several times. 

The query complexity of the algorithm \cite{abikpv2019} presented by Ambainis et.al is $O^*(1.728^n)$, where $O^*$ hides log-factors. The algorithm is based on the Grover's search algorithm which time complexity is more than the query complexity for additional log factor \cite{ad2017,g2002}.  So, the time complexity is also $O^*(1.728^n)$. If we want to discuss it more precisely, then we can check \cite{abikpv2019, kbcw2024} that claims the query complexity $O(n^31.728^n)$.

Finally, the complexity of the quantum algorithm for the problem is $O^*(1.728^n)$. At the same time, for current applications, we need only a classical solution. 
\subsection{Heuristic Solution of the Searching a Shortest Non-simple Path that Visits All Vertices Problem}\label{sec:heutisctic}

In the practical case, when we have more than 20-30 vertices in a graph, the exact algorithm that was presented before is too slow. In this section, we present a heuristic solution with polynomial time complexity, and the outcome is close, under certain criteria, to the optimal solution.

Using a path $P=(v_{i_1},\dots,v_{i_h})$, we can construct a sequence of edges $((v_{i_1},v_{i_2}),(v_{i_2},v_{i_3}),\dots,(v_{i_{h-1}},v_{i_h}))$. We can say that these are two equivalent representations of the path. We use both in this section.  

Note that the supergraph $S$ defined in Section \ref{sec:tools} satisfies the following conditions:
\begin{itemize}
\item $S$ is complete and weighted. 
\item the weight of an edge from $S$ is at most $n$ because it is the distance between two vertices in the original graph. Formally,  $w(v',u')=dist(v,u)\leq n$, where $v',u'\in V'$, and $v,u\in V$.
\item  $S$ is metric. It means that edges satisfy the triangle inequality: $w(v_i',v_j')\leq w(v_i',v_k')+w(v_k',v_j')$ for any $v_i',v_j',v_k' \in V'$. 
\end{itemize} 

Now we present a heuristic procedure $\textsc{2-Opt}(S)$ to find a good solution to the TSP problem. It finds a TSP cycle $C$. The algorithm for cycles in the case of general weighted metric graphs was presented in \cite{monnot2003approximation,paschos2016overview,hougardy2020approximation}. Here we present modifications for our graph and for searching a path.

We assume that we have two operations for modification of the list of edges that represents a cycle $C$:
\begin{itemize}
    \item $\textsc{Replace}(C, (v_{i_j},v_{i_{j+1}}), (u,v))$ that replaces the edge $(v_{i_j},v_{i_{j+1}})$ in $C$ by the edge $ (u,v)$ if $(v_{i_j},v_{i_{j+1}})$ is currently observed.
     \item $\textsc{Reverse}(C, (v_{i_j},v_{i_{j+1}}), (v_{i_k},v_{i_{k+1}}))$ that reverses all elements of the list between $(v_{i_j},v_{i_{j+1}})$ and $ (v_{i_k},v_{i_{k+1}})$. if $(v_{i_j},v_{i_{j+1}})$ and $ (v_{i_k},v_{i_{k+1}})$ are currently observed. Both elements $(v_{i_j},v_{i_{j+1}})$ and $ (v_{i_k},v_{i_{k+1}})$ keep their place.
\end{itemize}
Both operations can be implemented in $O(1)$ time complexity if $C$ is implemented by a Two-Directional Linked List data structure \cite{cormen2001}.

Firstly, we initialize the cycle by $((v_1,v_2),\dots,(v_{n-1},v_n),(v_n,v_1))$. Then we repeatedly replace two edges of the cycle as long as this yields a shorter cycle. We check all possible pairs. The implementation is presented in Algorithm \ref{alg:2opt}

\begin{algorithm}[H]
\caption{Implementation of $\textsc{2-Opt}(S)$}\label{alg:2opt}
\begin{algorithmic}
\State $C\gets ((v_1,v_2),\dots,(v_{n-1},v_n),(v_n,v_1))$
\State $stopFlag\gets False$
\While{$stopFlag=False$}
\State $stopFlag\gets True$
\For{$k \in \{1,\dots,n\}$}
\For{$j \in \{k+1,\dots, n\}$ }
\If{$w(v_{i_k},v_{i_{k+1}})+w(v_{i_j},v_{i_{j+1}}) > w(v_{i_{k}},v_{i_j})+w(v_{i_{k+1}},v_{i_{j+1}})$}
\State $\textsc{Reverse}(C,(v_{i_k},v_{i_{k+1}}), (v_{i_j},v_{i_{j+1}}))$
\State $\textsc{Replace}(C,(v_{i_k},v_{i_{k+1}}), (v_{i_k},v_{i_{j+1}}) )$
\State $\textsc{Replace}(C,(v_{i_j},v_{i_{j+1}}), (v_{i_{k+1}},v_{i_{j}}))$
\State $stopFlag\gets False$
\State break both For-loops
\EndIf
\EndFor
\EndFor
\EndWhile
\State \Return $C$
\end{algorithmic}
\end{algorithm}

In the case of searching for a path, we have two modifications. Namely, if a starting vertex $v_{s}$ is fixed, then we initialize the path by 
\[C\gets ((v_{s},v_1),(v_1,v_2),\dots,(v_{s-2},v_{s-1}),(v_{s+1},v_{s+2}),\dots(v_{n-1},v_n)).\]
If the starting vertex is not fixed, then we check all possible vertices from $V'$ as the starting vertex $v_s$. 

 Experiments on real-world instances have shown that $\textsc{2-Opt}(S)$ achieves much better results than Christofide's algorithm \cite{bentley1992fast}. Generally, the worst case of $\textsc{2-Opt}(S)$ may lead to exponential complexity. However, the direct analysis shows that the complexity of $\textsc{2-Opt}(S)$ in the case of supergraphs is $O(n^4)$ for cycles and for paths with fixed starting vertex, where $n$ is the number of vertices.
\begin{theorem}
The time complexity of $\textsc{2-Opt}(S)$ for supergraph $S$ constructed by $G$ is $O(n^4)$ in the case of a cycle and a path with a fixed starting vertex, and $O(n^5)$ in the case of a path.
\end{theorem}
\Beginproof
Initialization of $C$ takes $O(n)$ time. Since weights of edges are bounded above by $n$, the weight of any Hamiltonian cycle $C$ is bounded above by $n^2$. In each step of ``replacing two edges'', the weight of $C$ reduces at least to $1$, which implies that there are $O(n^2)$ steps of ``replacing two edges''. In each step, there are $O(n^2)$ pairs of edges to compare, and comparison and replacement of edges take constant time. In summary, the total complexity is $O(n^4)$.

In the case of a path, we also check all starting vertices. The number of starting vertices is $n$. So, the complexity is $O(n^5)$ for searching the path.
\Endproof

The heuristic algorithm $\textsc{2-Opt}(S)$ is designed to compute feasible solutions that are as close as possible to the optimal ones with respect to some criterion. We present two paradigms dealing with polynomial approximation. More details can be found in \cite{ausiello2012complexity,paschos2016overview}. 

\begin{definition}
     Let $I$ be an input, and ${\cal A}$ be an  approximation algorithm. Let $len_{opt}(I)$ be the length of a path that returns an optimal solution for the given instance $I$, $len_{w}(I)$ be the worst possible length that a feasible solution returns for $I$, and $len_{{\cal A}}(I)$ be the length of the path that returns the approximation algorithm ${\cal A}$. Then, For an istance $I$
     \begin{itemize}
         \item the standard approximation ratio is
         
         $\rho_{{\cal A}}(I)=\frac{len_{{\cal A}}(I) }{ len_{opt}(I)}$;
         \item the differential approximation ratio is 
         
         $\delta_{{\cal A}}(I)=\frac{|len_{w}(I)-len_{{\cal A}}(I)|}{|len_w(I)-len_{opt}(I)|}$.
     \end{itemize}
\end{definition}
Note that if the ratios are close to $1$, then the algorithm has better quality.

For the heuristic algorithms of the traveling salesman problem, we have the following results.

\begin{lemma}[\cite{monnot2003approximation,paschos2016overview,hougardy2020approximation}]
    The differential approximation ratio for $\textsc{2-Opt}(S)$ solution of the Traveling salesman problem achieves $\frac{1}{2}$, and if the graph is metric, the standard approximation ratio achieves $\sqrt{\frac{n}{2}}$. Both of the two ratios are tight. 
\end{lemma}

\section{Method for Constructing a Circuit for Quantum Hashing}\label{sec:hash}

\begin{figure}[H]
\includegraphics[width=0.4\textwidth]{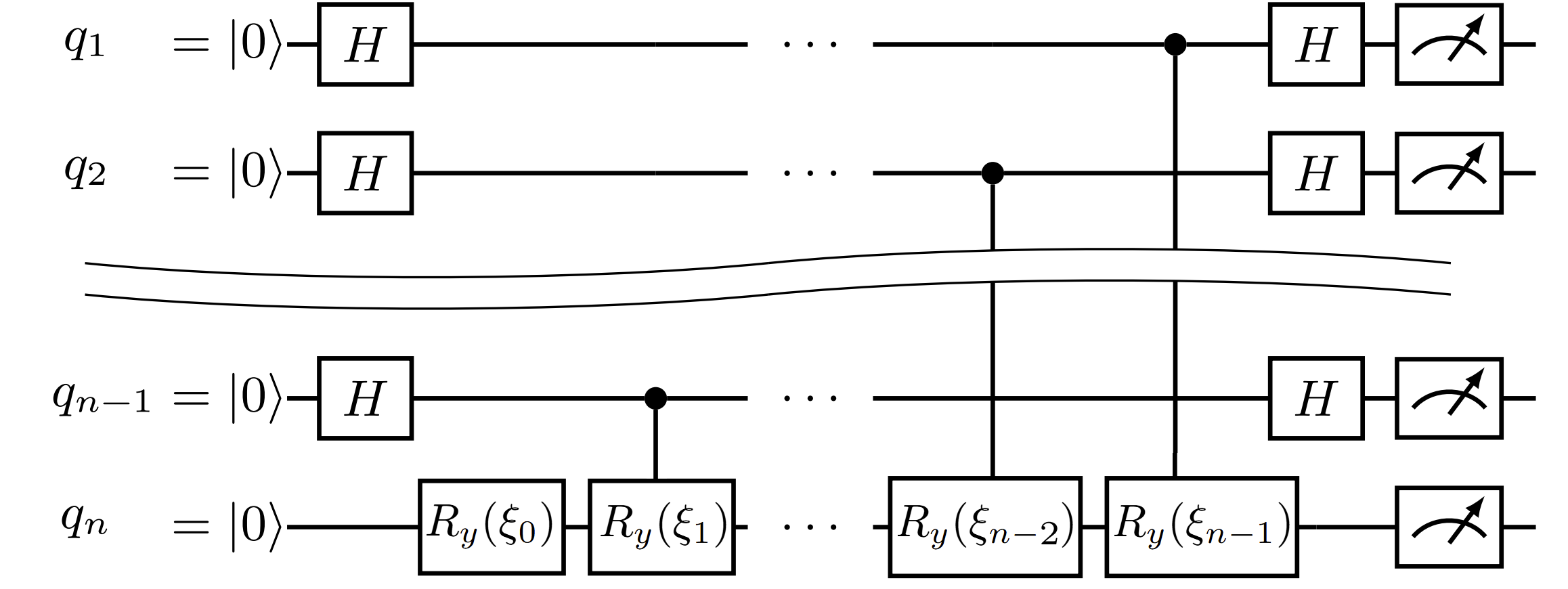}
\caption{\label{fig:qf} Shallow circuit for quantum fingerprinting or quantum hashing algorithm}
\end{figure}

In this section, we present a method that allows us to construct a circuit for the quantum fingerprinting or quantum hashing algorithm for a connected graph $G=(V,E)$ that is qubits connection graph for a device. More information about the quantum fingerprinting (quantum hashing) algorithm can be found in Appendix \ref{apx:hash}.  Here we consider a shallow circuit \cite{kalis18,ziiatdinov2023gaps} for $n$ control qubits. If we do not have restrictions for applying two-qubit gates (when $G$ is a complete graph or a ``star'' graph), then the circuit is such that we present in Figure \ref{fig:qf}. Here we assume that we have qubits $q_1,\dots,q_{n}$ such that $q_1,\dots,q_{n-1}$ are control ones and $q_{n}$ is the target one.

The logical qubits from the original circuit we call $q_i$ and the qubits of a physical device are associated with vertices of the graph $G$, and we call them $v_i$..

The algorithm for constructing a circuit is the following.

\begin{itemize}
    \item[] \textbf{Step 1.} We find the shortest non-simple path in the graph $G$ that visits all vertices at least once using the algorithm from Section \ref{sec:tools}. Assume that the path is $P=(v_{i_1}, \dots,v_{i_k})$.
    \item[]  \textbf{Step 2.} The target qubit $q_{n}$ corresponds to the vertex $v_{i_2}$. Control qubits $q_1,\dots,q_{n-1}$ correspond to other vertices.  We assume that we have a set $U$ of control qubits that are already used. Initially, it is empty $U\gets\emptyset$. Let $j$ be the index of the element in the path $P$ that corresponds to the target qubit. Initially, $j\gets 2$.
    \item[]  \textbf{Step 3.} We apply control rotation with the control $v_{i_1}$ and the target $v_{i_2}$ qubits. Then, we add $v_{i_1}$ to the set $U$, i.e. $U\gets U\cup\{v_{i_1}\}$.
    \item[]  In the next steps the target qubit travels by the path $P$. 
    \item[]  \textbf{Step 4.} If $v_{i_{j+1}}\not\in U$, then we apply a control rotation to the control $v_{i_{j+1}}$ and the target $v_{i_j}$ qubits. After that, we add $v_{i_{j+1}}$ to the set $U$, i.e. $U\gets U\cup\{v_{i_{j+1}}\}$. If $j=n-1$, then we terminate the algorithm. Otherwise, we go to Step 5.
     \item[]  \textbf{Step 5.} If $v_{i_{j+2}}= v_{i_{j}}$, then we go to Step 6. Otherwise, we apply SWAP gate to $v_{i_{j}}$ and $v_{i_{j+1}}$. After that, we update $j\gets j+1$ because the value of the target qubit moves to $v_{i_{j+1}}$. Then, we go to Step 4.
     \item[] \textbf{Step 6.} If $v_{i_{j+2}}=v_{i_{j}}$, then we update $j\gets j+2$. Note that here we stay on the same vertex of the graph $G$. Then, we go to Step 4.     
\end{itemize}

Let us present a procedure $\textsc{ConsrtuctForPath}(P)$ in Algorithm \ref{alg:cascade} that implements the above idea for a given path $P=(v_{i_1}, \dots,v_{i_k})$. We assume that we have $\textsc{cR}(u,v)$ procedure that applies the control rotation operator $CR_y$ to $u$ as a control qubit and $v$ as a target one. The angle $\xi_r$ corresponds to the qubit $q_r$ associated with the vertex $v$. Additionally, we have $\textsc{swap}(u,v)$ procedure that applies swap gate to $u$ and $v$ qubits.

\begin{algorithm}[H]
\caption{Implementation of $\textsc{ConsrtuctForPath}(P)$ procedure. Algorithm of constructing circuit for quantum hashing or quantum fingerprinting for a path $P=(v_{i_1}, \dots,v_{i_k})$}\label{alg:cascade}
\begin{algorithmic}
\State $U\gets\emptyset$
\State $j\gets 2$
\State $\textsc{cR}(v_{i_1},v_{i_2})$
\State $U\gets U\cup\{v_{i_1}\}$
\State  $\textsc{cR}(v_{i_{j+1}},v_{i_j})$
\State $U\gets U\cup\{v_{i_{j+1}}\}$
\While{$j< k-1$}
\If{$v_{i_{j+2}}= v_{i_{j}}$}
\State $j\gets j+2$
\Else
\State $\textsc{swap}(v_{j},v_{j+1})$
\State $j\gets j+1$
\EndIf
\If{$v_{j+1}\not\in U$}
\State  $\textsc{cR}(v_{i_{j+1}},v_{i_j})$
\State $U\gets U\cup\{v_{i_{j+1}}\}$
\EndIf
\EndWhile
\end{algorithmic}
\end{algorithm}
We assume that $\textsc{ShortestNSPath}(G)$ returns a shortest non-simple path in the graph $G$ that visits all vertices at least once. Implementation of this procedure can be found in Algorithm \ref{alg:tsp} or Algorithm \ref{alg:2opt}. Finally, we can present Algorithm \ref{alg:qh} as an implementation of the quantum fingerprinting (quantum hashing) algorithm.
\begin{algorithm}[H]
\caption{Algorithm of constructing circuit for quantum hashing or quantum fingerprinting for a path $P=(v_{i_1}, \dots,v_{i_k})$}\label{alg:qh}
\begin{algorithmic}
\State $P=(v_{i_1}, \dots,v_{i_k})\gets \textsc{ShortestNSPath}(G)$
\State  $\textsc{ConsrtuctForPath}(P)$ 
\end{algorithmic}
\end{algorithm}

\begin{figure}[H]
\includegraphics[width=0.4\textwidth]{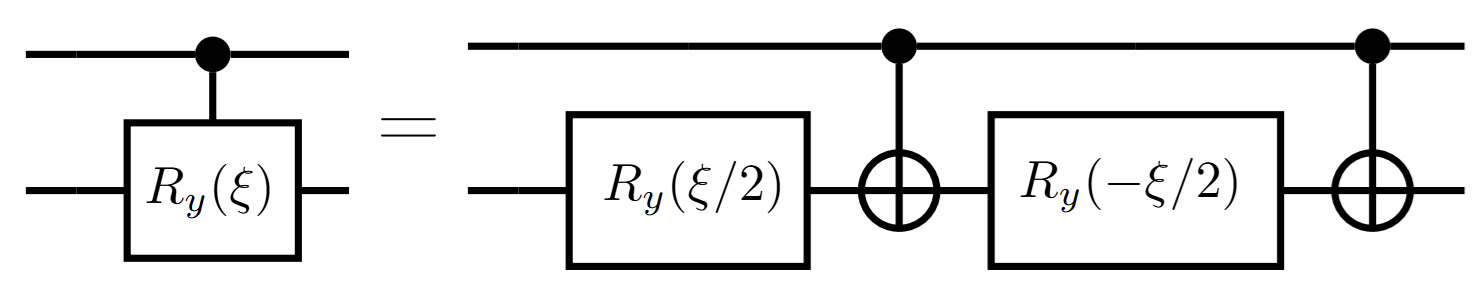}
\caption{\label{fig:cr} Representation of $CR_y$ gate using only basic gates}
\end{figure}

We can represent the control rotation operator $\textsc{cR}(u,v)$ and the corresponding angle $\xi$ as a sequence of $\textsc{R}(v,\xi/2)$, $\textsc{cnot}(u,v)$, $\textsc{R}(v,-\xi/2)$, and $\textsc{cnot}(u,v)$, (see Figure \ref{fig:cr}). Here $\textsc{R}(v,\xi/2)$ is the $R_y(\xi/2)$ gate applied to the qubit $v$; $\textsc{cnot}(u,v)$ is the $CNOT$ for $u$ as a control and $v$ as a target.

\begin{figure}[H]
\includegraphics[width=0.27\textwidth]{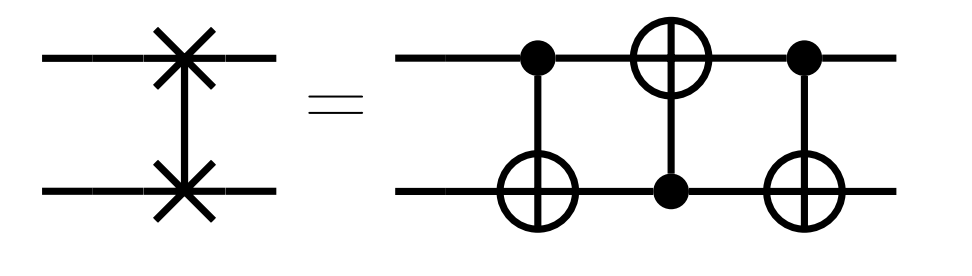}
\caption{\label{fig:swap} Representation of $SWAP$ gate using only basic gates}
\end{figure}

Additionally, the  $\textsc{swap}(u,v)$  gate can be represented as a sequence $\textsc{cnot}(u,v)$, $\textsc{cnot}(v,u)$, and $\textsc{cnot}(u,v)$ (see Figure \ref{fig:swap}).

\begin{figure}[H]
\includegraphics[width=0.4\textwidth]{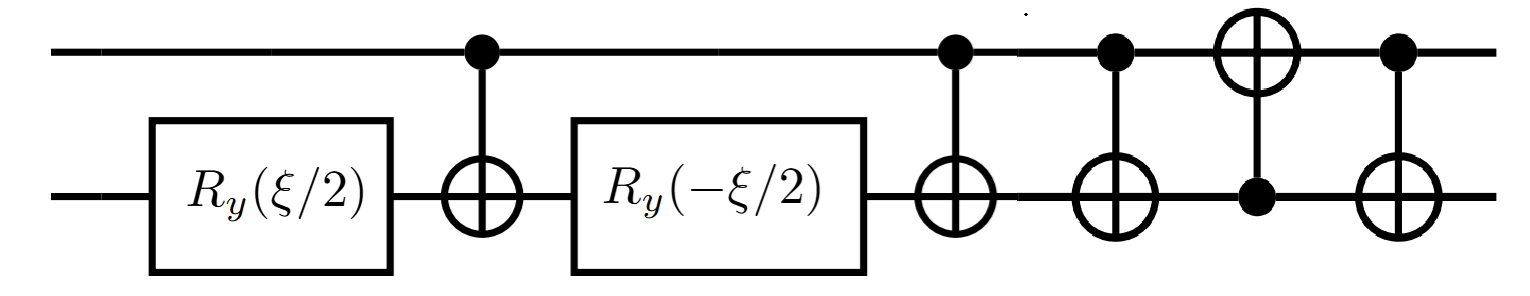}
\caption{\label{fig:crswap1} Representation of a pair $CR_y$ and $SWAP$ gates using only basic gates}
\end{figure}

Two sequential operators $CR_y$ and $SWAP$ that are $\textsc{cR}(u,v)$ and  $\textsc{swap}(u,v)$ procedures can be represented by a circuit in Figure \ref{fig:crswap1}.

\begin{figure}[H]
\includegraphics[width=0.35\textwidth]{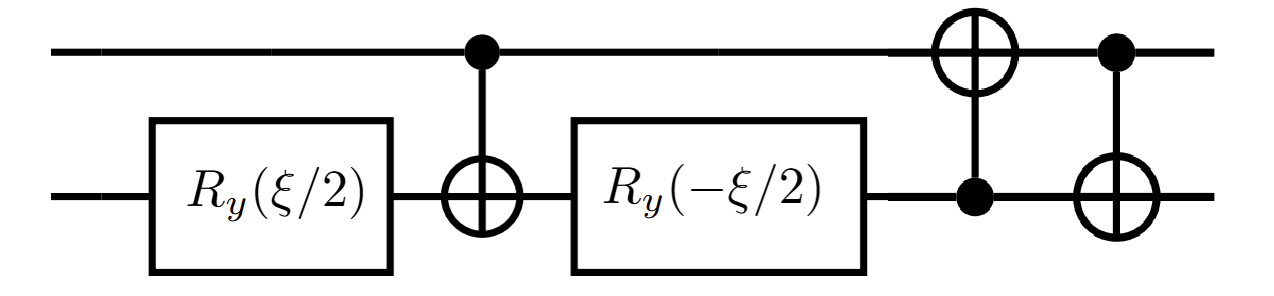}
\caption{\label{fig:crswap2} Reduced representation of a pair $CR_y$ and $SWAP$ gates using only basic gates}
\end{figure}

Note that two sequential $\textsc{cnot}(u,v)$ gates are annihilated,  as presented in \cite{k2024aliya,ksy2024}. Thise we obtain the circuit in Figure \ref{fig:crswap2}.

Let us look at the CNOT cost of these operators, where CNOT cost is the number of CNOT operators in the representation of a circuit using only basic gates.

We can say that CNOT cost of $\textsc{cR}(u,v)$ is $2$; CNOT cost of $\textsc{swap}(u,v)$ is $3$, CNOT cost of two sequential operators $\textsc{cR}(u,v)$ and  $\textsc{cnot}(u,v)$ is $3$. 

Finally, we can discuss the CNOT cost of the constructed circuit.
\begin{theorem}\label{th:qh1}
    CNOT cost of the circuit that is generated by Algorithm \ref{alg:qh} is $3k-b-2\leq 3k-2$, where $k$ is the length of a shortest non-simple path $P=(v_{i_1},\dots,v_{i_k})$ in the graph $G$ that visits all vertices at least once, and $b$ is the number of indexes $j$ such that $v_{i_j}=v_{i_{j+2}}$.  
\end{theorem}
\Beginproof
If we look at Algorithm \ref{alg:qh}, then we can see that it is a sequence of one of three options:
\begin{itemize}
    \item a pair of $CR_y$ and $SWAP$ gates if $v_{j}\neq v_{j+2}$ and $v_{j+1}\not\in U$.
    \item  $SWAP$ gate if $v_{j}\neq v_{j+2}$ and $v_{j+1}\in U$.
    \item $CR_y$ gate if $v_{j}= v_{j+2}$ and $v_{j+1}\not\in U$.
\end{itemize}
Note that the option $v_{j}= v_{j+2}$ and  $v_{j+1}\in U$ is not possible because  $v_{j+1}\in U$ means that the corresponding qubit already visited. In that case, we can just exclude it from the path because $v_{j}= v_{j+2}$. At the same time, it contradicts the condition that $P$ is the shortest path.

The CNOT cost for pairs of $CR_y$ and $SWAP$ gates is $3$; for the $SWAP$ gate, it is $3$; for the $CR_y$ gate, it is $2$.

Let the number of steps where we apply pairs of $CR_y$ and $SWAP$ gates or the $SWAP$ gate be $a$, and the number of steps where we apply  $CRy$  be $b$. Here $a+b=k-2$ because the first and the last steps are $CRy$ gates without SWAP gate, and each of them costs $2$. So, the total number of CNOT gates is $3a+2b + 4= 3(a+b+2) -b - 2=3k - b-2$.
\Endproof

Using Lemma \ref{lm:path-len} we can estimate the CNOT cost of the circuit, and present it in the next corollary.

\begin{corollary}\label{cr:path}
CNOT cost of the circuit that is generated using Algorithm \ref{alg:qh} is in range from $3n-2$ to $\frac{3}{2}n^2 +\frac{3}{2}n-2$. 
\end{corollary}

Typically, the quantum fingerprinting (quantum hashing) algorithm is used several times. For example, for hashing a string of $\ell$ symbols we should apply the algorithm $\ell$ times (see Appendix \ref{apx:hash} for examples). For this reason, we suggest two strategies.

\paragraph{Strategy with a Path.} Let a step be a single application of the quantum fingerprinting algorithm.
After each step, we reverse the path $P$. Then, the target qubit travels back. Additionally, we can see that the operator $\textsc{cR}(v_{i_{k_1}},v_{i_k})$ 
is applied at the end of the odd step and at the beginning of the even step. We can apply it once with a double angle. The same situation with  $\textsc{cR}(v_{i_{1}},v_{i_{2}})$ on the meeting of even and odd step. Therefore, we have the following CNOT cost for $\ell$ applications of the quantum hashing algorithm.

\begin{theorem}
CNOT cost of the circuit for application of the quantum fingerprinting (quantum hashing) algorithm $\ell$ times is at most $(3k-4)\ell+2$, where  $k$ is the length of the shortest non-simple path $P=(v_{i_1},\dots,v_{i_k})$ in the graph $G$ that visits all vertices at least once.
\end{theorem}
\Beginproof
The first step costs $3k-2$ due to Theorem \ref{th:qh1}. Each next step skips the first $\textsc{cR}$ gate because it is ``merged'' with the last $\textsc{cR}$ from the previous step. So, its CNOT cost is $3k-4$ because $\textsc{cR}$ costs $2$. The total cost is $3k-2 + (3k-4)(\ell-1)=(3k-4)\ell+2$.
\Endproof

Using Lemma \ref{lm:path-len} we can estimate the CNOT cost of the circuit in the next corollary.

\begin{corollary}\label{cr:cycle2}
CNOT cost of the circuit for application of the quantum fingerprinting (quantum hashing) algorithm $\ell$ times is in the range from $(3n-4)\ell+2$ to $(\frac{3}{2}n^2+\frac{3}{2}n-4)\ell+2$. 
\end{corollary}

\paragraph{Strategy with a Cycle.} At the beginning of each step starting from the second one we add $\textsc{swap}(v_{i_k},v_{i_1})$ and $\textsc{swap}(v_{i_1},v_{i_2})$ gates. We need the first $\textsc{swap}$ because we should start from the beginning of the cycle next time. We add the second one because we should start from the second element of the cycle. These two gates increase the cost by $2$ because they are joint with corresponding $\textsc{cR}$ gates. At the same time, all logical qubits are moved to $1$ position of the cycle. That is why we need only $k-1$ travels by the cycle for doing $k$ steps.

Therefore, we have the following CNOT cost for $\ell$ applications of the quantum hashing algorithm.
\begin{theorem}
CNOT cost of the circuit for application of the quantum fingerprinting (quantum hashing) algorithm $\ell$ times is at most $(3k-3)(\ell+1)+1$, where  $k+1$ is the length of the shortest non-simple cycle $C=(v_{i_1},\dots,v_{i_k},v_{i_1})$ in the graph $G$ that visits all vertices at least once.
\end{theorem}
\Beginproof
The first step costs $3k-2$ due to Theorem \ref{th:qh1}. Each next step costs $3k$ because of two additional $\textsc{swap}$ gates. At the same time, we should apply them only $\ell-\left\lfloor\frac{\ell}{k}\right\rfloor$ because of the moving of the logic qubits by the cycle.

The total cost is $3k-2 + 3k(\ell-\left\lfloor\frac{\ell}{k}\right\rfloor-1)=
3k-2 + 3k\ell-3k\left\lfloor\frac{\ell}{k}\right\rfloor-3k\leq
3k-2 + 3k\ell-3k\frac{\ell}{k}+3k-3k=
3k-2 + 3k\ell-3\ell=
3(k-1)+1 + 3(k-1)\ell
=(3k-3)(\ell+1)+1$.
\Endproof

Using Lemma \ref{lm:path-len} we can estimate the CNOT cost of the circuit in the next corollary.

\begin{corollary}\label{cr:cycle}
CNOT cost of the circuit for application of the quantum hashing (quantum fingerprinting) algorithm $\ell$ times is in the range from $(3n-3)(\ell+1)+1$ to $(\frac{3}{2}n^2+\frac{3}{2}n-3)(\ell+1)+1$. 
\end{corollary}

We can see that the complexity presented in Corollary \ref{cr:path} is less than the result from Corollary \ref{cr:cycle} because each cycle that visits all the vertices at least once can be converted to the path by removing the last edge. At the same time, the repeated application of the same pattern of gates for each step can be useful in some hardware implementations.

\subsection{Comparing with Existing Results}\label{sec:compare-qh}
Let us compare our generic approach with existing circuits for specific graphs. The LNN architecture is a common layout graph for many quantum devices. The graph is a chain where $v_1$ is connected with $v_2$;  $v_i$ is connected with $v_{i-1}$ and $v_{i+1}$ for $i\in\{2,\dots,n-1\}$; $v_n$ is connected with $v_{n-1}$.

\begin{figure}[H]
\begin{center}    \includegraphics[width=0.3\textwidth]{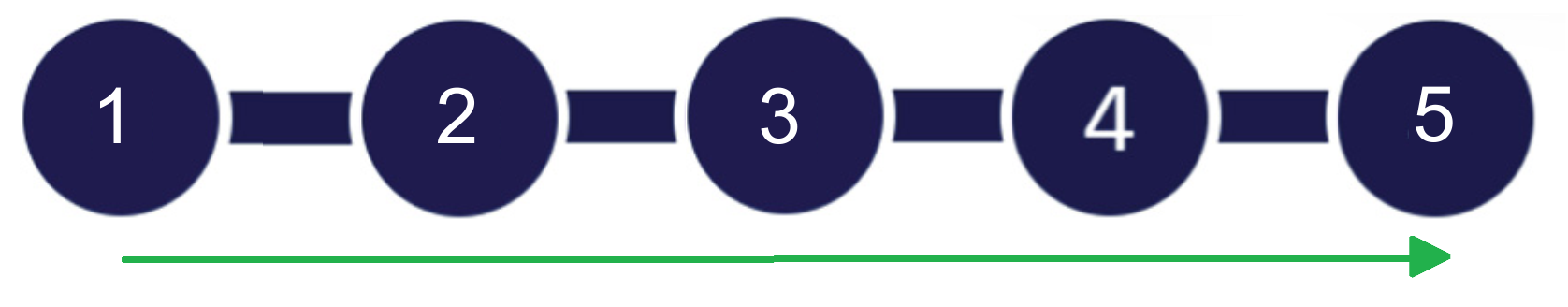}
\end{center}
\caption{\label{fig:lnn-5} The graph for the 5-qubit LNN architecture. The path that visits all vertices at least once is green.}
\end{figure}

\begin{figure}[H]
\includegraphics[width=0.5\textwidth]{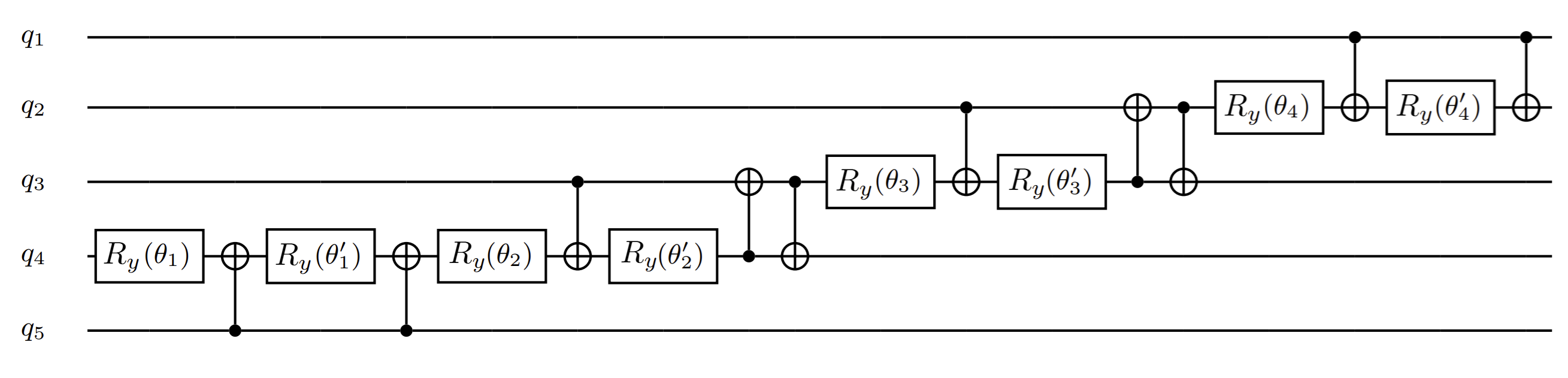}
\caption{\label{fig:lnn-qhash} A quantum circuit for Quantum hashing (quantum fingerprinting) algorithm for $5$ qubits LNN architecture device}
\end{figure}

The quantum fingerprinting (quantum hashing) algorithm for the LNN architecture was discussed in \cite{ksy2024}. The path that visits all vertices in the graph is simply $P=(v_1,\dots,v_n)$ (see Figure \ref{fig:lnn-5}).
So, our method gives us the same result circuit as in  \cite{ksy2024}. The circuit for the quantum fingerprinting algorithm on $5$ qubits is presented in Figure \ref{fig:lnn-qhash}.
The CNOT cost of the circuit is presented in the next lemma
\begin{lemma}
The CNOT cost of the produced circuit of the quantum fingerprinting (quantum hashing) algorithm with $n$ qubits for the LNN architecture is $3n-5$ for one application and $3n\ell-7\ell +2$ for $\ell$ applications.
\end{lemma}

\begin{figure}[H]
\begin{center}
\includegraphics[width=0.3\textwidth]{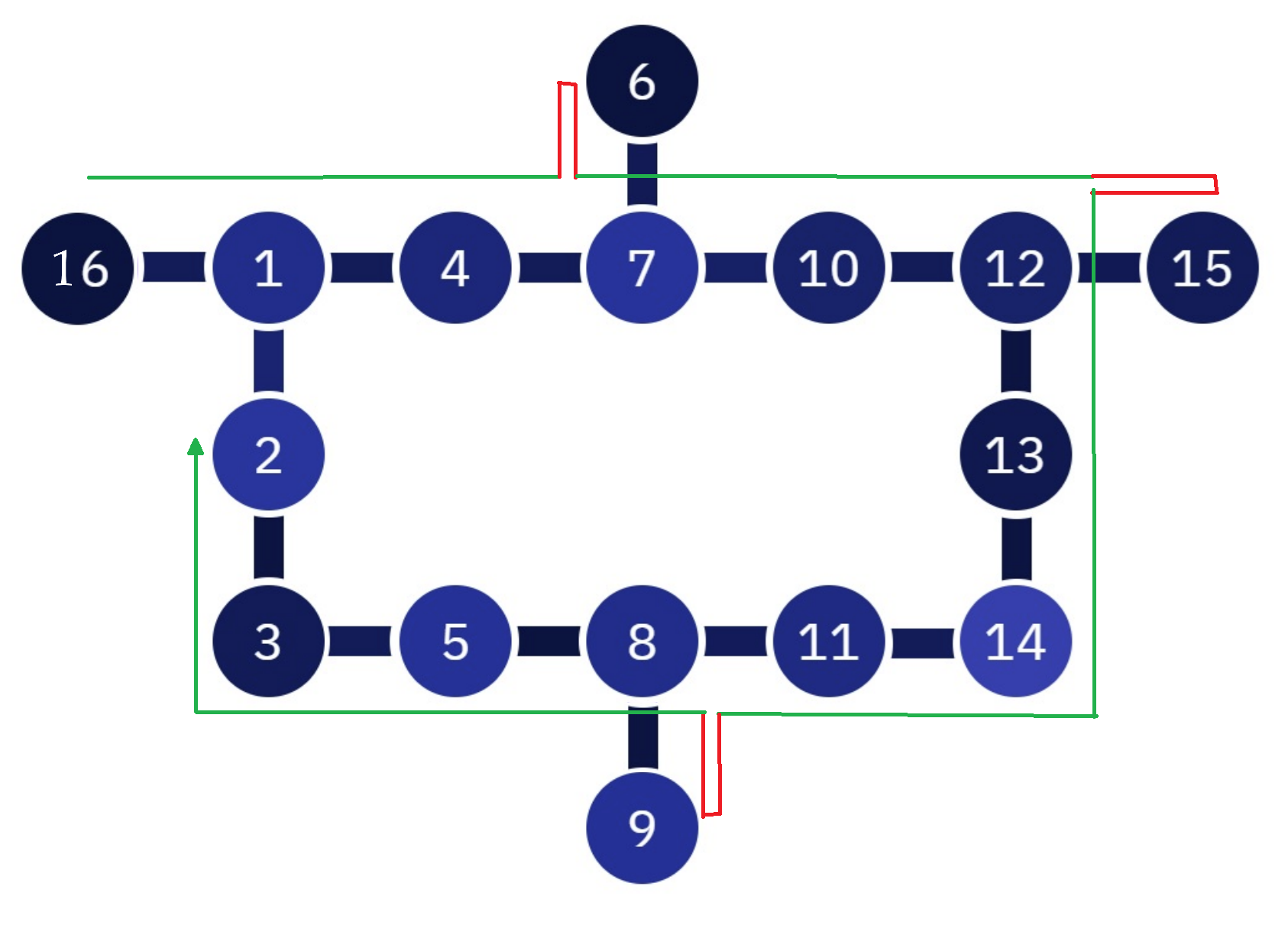}
\end{center}
\caption{\label{fig:sun1} 
The graph for 16-qubit ``sun'' architecture. The path that visits all vertices at least once is green. Red parts are such that $v_{i_j}=v_{i_{j+2}}$}
\end{figure}

\begin{figure}[H]
\includegraphics[width=0.45\textwidth]{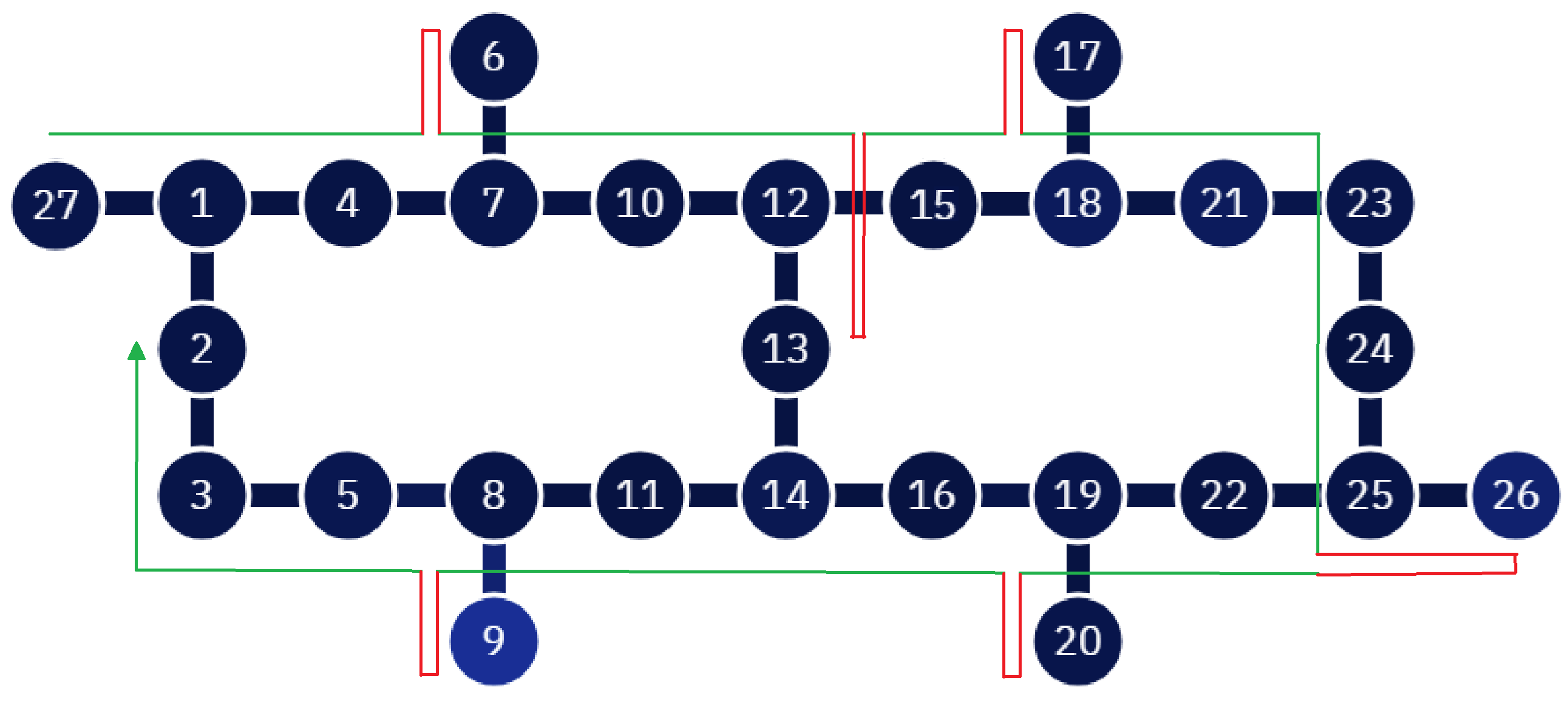}
\caption{\label{fig:cairo1} 
The graph for 27-qubit ``two joint suns'' architecture. The path that visits all vertices at least once is green. Red parts are such that $v_{i_j}=v_{i_{j+2}}$}
\end{figure}

Let us consider more complex graphs like a cycle with tails (like a ``sun'' or ``two joint suns''). The 16-qubit and 27-qubit IBM machines of such architecture were considered in \cite{k2024aliya}. The graphs and paths are 
presented in Figures \ref{fig:sun1} and \ref{fig:cairo1}.


Note that the red part of the path is such that $v_{j+2}=v_{j}$. Therefore, we do not apply a SWAP gate on this part of the path, but only a $CR_y$ gate. So, our generic method gives the same circuits as the circuits specially constructed for these devices \cite{k2024aliya}. The CNOT cost is $42$ for 16-qubit architecture, and $69$ for 27-qubit one. 


\section{Method for Constructing a Circuit for Quantum Fourier Transform}\label{sec:qft}

\begin{figure}[H]
\includegraphics[width=0.5\textwidth]{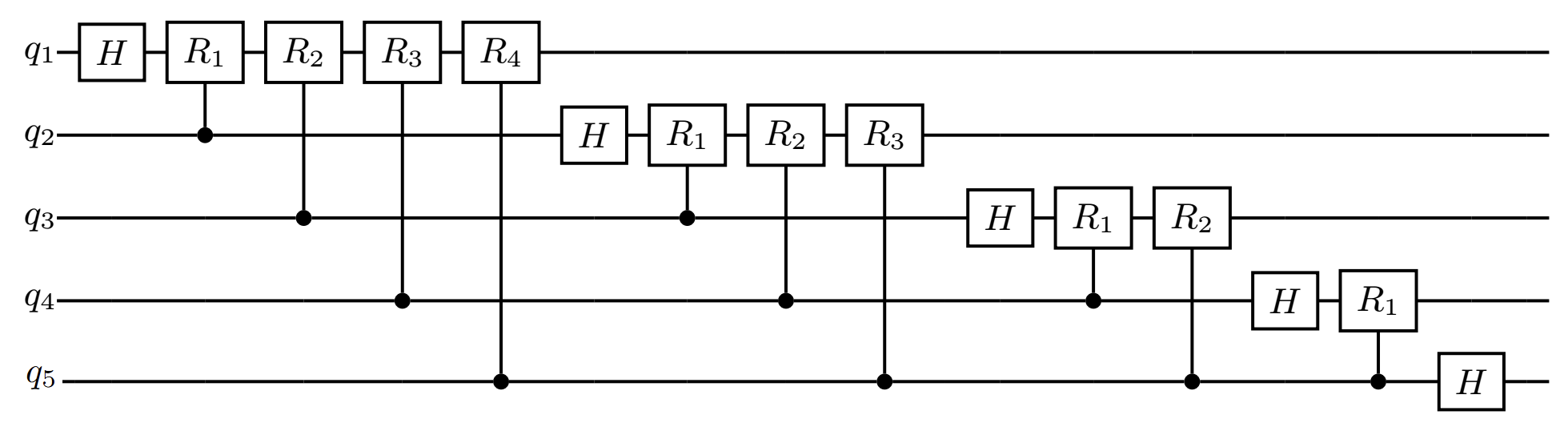}
\caption{\label{fig:cqft} A quantum circuit for Quantum Fourier Transform algorithm for fully connected $5$ qubits}
\end{figure}

Here we present a method that allows us to construct a circuit for Quantum Fourier Transform (QFT) for a connected graph $G=(V,E)$ that is qubits connection graph for a device. More information about the QFT algorithm can be found in Appendix \ref{apx:qft}. If we do not have restrictions for applying two-qubit gates (when $G$ is a complete graph for instance), then the circuit is such as presented in Figure \ref{fig:cqft}.

The author of \cite{k2024aliya} show that the circuit for QFT is a series of CNOT cascades. Each of cascades has a gate structure similar to the shallow circuit for the quantum hashing algorithm. Assume that we have a $\textsc{ConsrtuctQFTForPath}(P',k',r)$ procedure that constructs the $r$-th cascade of the circuit for QFT for a path $P'$ of length $k'$. Note that for QFT, the size of each cascade is different. Here $P'$ is a path that visits only vertices corresponding to the qubits used in the current cascade. Firstly, we present the main algorithm in Section \ref{sec:qft1}. Then we present an algorithm for the $\textsc{ConsrtuctQFTForPath}(P',k',r)$ procedure in Section \ref{sec:qft2}. After that we discuss the complexity of the circuit in Section \ref{sec:qft3}. Finally, we compare the circuit with existing results in Section \ref{sec:qft4}. 

\subsection{The Main Algorithm}\label{sec:qft1}
Let us present the whole algorithm for constructing the quantum circuit for the QFT algorithm. 

Firstly, we find a non-simple path $P=(v_{i_1},\dots,v_{i_k})$ that visits all vertices at least once. Then we assign logical qubits to vertices according to this path. Let indexes $j_1,\dots,j_n$ be the indexes of the first occurrence of each vertex. In other words, $j_r$ is such that $i_{j_r}=r$, and $i_h\neq r$ for each $1\leq h<j_r$.

Then, we assign $Q_r\gets j_r$ the position of $r$-th logical qubit in the graph, and $T_{j_r}\gets r$ the index of logical qubit associated with a vertex.

After that, we start to construct the circuit for cascades. For the first cascade, we invoke $\textsc{ConsrtuctQFTForPath}(P,k)$.
Similarly to the quantum hashing (quantum fingerprinting) algorithm, the target $1$-st logical qubit moves to the end of the path, i.e. $Q_1=i_k$.

For the second cascade, we exclude the vertex with an index $Q_1$ from the graph and find the path $P'$ that visits all vertices at least once in the updated graph such that it starts from the vertex $v_{Q_2}$. The vertex $v_{Q_2}$ contains the $2$-nd logical qubit that should be the target in the second cascade. Then, we invoke $\textsc{ConsrtuctQFTForPath}(P',k')$, where $k'$ is the length of $P'$. 

For the next cascades, we act in a similar way. For the $r$-th cascade, we exclude $v_{Q_{r-1}}$, find the path $P'$ that visits all vertices at least once in the updated graph such that it starts from the vertex $v_{Q_r}$, and invoke $\textsc{ConsrtuctQFTForPath}(P',k')$.

Note that a path in the updated graph always exists because it is connected. We have at least one path that visits all vertices at least once, and we exclude only one vertex that is the last one on this path.

During the implementation, we do not remove the vertex from the graph, we just store a set of ``deleted'' vertices and we prohibit visiting vertices from this set during the algorithm for the path search.

Let us present the algorithm.

\begin{itemize}
 \item[]\textbf{Step 1}. We find a non-simple path $P=(v_{i_1},\dots,v_{i_k})$ that visits all vertices at least once.
 
 \item[]\textbf{Step 2}. We find indexes $j_1,\dots,j_n$, where $j_r$ is such that $i_h\neq i_{j_r}$ for each $1\leq h<j_r$. Then, we assign $Q_r\gets i_{j_r}$ and $T_{i_{j_r}}\gets r$ for $1\leq r\leq n$.

  \item[]\textbf{Step 3}. We assign $P'\gets P$, and $k'$ is the length of $P'$. We assign $r\gets 1$ which is an index of the cascade.
\end{itemize}
The next steps are repeated until the graph contains at least one vertex.
\begin{itemize}    

    \item[]\textbf{Step 4}. We find a non-simple path $P'$ of length $k'$ that visits all non-deleted vertices of the graph and starts from $v_{Q_r}$. If $r=1$, then we do nothing because $P'$ was already found in Steps 1-3. 
    \item[]\textbf{Step 5}. We invoke $\textsc{ConsrtuctQFTForPath}(P',k',r)$ that is the $r$-th cascade. Assume that the procedure keeps the $Q$ and $T$ indexes actual.
    \item[]\textbf{Step 6}. We exclude $v_{Q_r}$ from the graph.
    \item[]\textbf{Step 7}. If $r=n$, then we terminate the algorithm. Otherwise, we increase the index of the cascade $r\gets r+1$ and go to Step 4.    
\end{itemize}
The implementation of the algorithm is presented in Algorithm \ref{alg:qft-whole}. Assume that we have the $\textsc{ShortestNSPath}(G)$ procedure that returns the shortest non-simple path visiting all vertices, and 
 the $\textsc{ShortestNSPath}(G,v)$ procedure that returns a similar path that starts from the vertex $v$. Let $\textsc{getFirstIndexes}(P)$ be a procedure that returns $j_1,\dots,j_n$. The implementation of this procedure is simple, has $O(k\log n)$ time complexity, and is presented in Appendix \ref{apx:firstindexes}. 
\begin{algorithm}[H]
\caption{Implementation of the algorithm of constructing the whole circuit for QFT for a path $P=(v_{i_1}, \dots,v_{i_k})$}\label{alg:qft-whole}
\begin{algorithmic}
\State $P\gets \textsc{ShortestNSPath}(G)$, $k\gets len(P)$
\State $(j_1,\dots,j_n)\gets \textsc{getFirstIndexes}(P)$
\For{$r\in\{1,\dots,n\}$}
    \State $Q_r\gets i_{j_r}$
    \State $T_{i_{j_r}}\gets r$
\EndFor
\State $P'\gets P$, $k'\gets k$, $r\gets 1$
\While{$r\leq n$}
    \State $\textsc{ConsrtuctQFTForPath}(P',k',r)$
    \State $V\gets V\backslash \{v_{Q_r}\}$
    \State $r\gets r+1$
    \If{$r\leq n$}
        \State $P'\gets \textsc{ShortestNSPath}(G,v_{Q_r})$, $k'\gets len(P')$
    \EndIf
\EndWhile
\end{algorithmic}
\end{algorithm}

Let us discuss the time complexity of the algorithm. The time complexity of $\textsc{ShortestNSPath}(G)$ is $O(n^2 2^n)$, and the complexity of $\textsc{ShortestNSPath}(G,v)$ is at most $O(n 2^n)$ due to Theorem \ref{th:path-compl}. So, the total complexity is $O(n^2 2^n)$. In the case of the heuristic solution, we have a similar situation. The complexity of the presented solution from Section \ref{sec:heutisctic} is $O(n^5)$.

\subsection{Quantum Circuit for One Cascade}\label{sec:qft2}

The procedure $\textsc{ConsrtuctQFTForPath}(P',k',r)$ is very similar to the $\textsc{ConsrtuctForPath}(P)$ procedure that constructs the circuit for the quantum hashing algorithm. At the same time, there are a lot of small specific modifications. That is why we present the whole algorithm here.
\begin{itemize}
    \item[] \textbf{Step 1.} We start with the first qubit in the path $j\gets 1$. We apply the Hadamard transformation to the qubit corresponding to the vertex $v_{i_1}$. We denote this action by $\textsc{H}(v_{i_1})$. If $k'=1$, then we terminate our algorithm; otherwise, go to Step 2. 
        \item[]  \textbf{Step 2.} If $v_{i_{j+1}}\not\in U$, then we apply the control phase gate $CR_d$ with the control $v_{i_{j+1}}$ and the target $v_{i_j}$ qubits, where $d=T_{i_{j+1}}-r$. Then, we add $v_{i_{j+1}}$ to the set $U$, i.e. $U\gets U\cup\{v_{i_{j+1}}\}$.  If $j=k-1$, then we terminate the algorithm. Otherwise, we go to Step 3.
     \item[]  \textbf{Step 3.} If $v_{i_{j+2}}= v_{i_{j}}$, then we go to Step 4. Otherwise, we apply the SWAP gate to $v_{i_{j}}$ and $v_{i_{j+1}}$, and swap the indexes of qubits for these vertices. In other words, if $w_1=T_{i_j}$ and $w_2=T_{i_{j+1}}$ are indexes of the corresponding logical qubits, then we swap $Q_{w_1}$ and $Q_{w_2}$ values, and $T_{i_j}$ and $T_{i_{j+1}}$ values. Then, we update $j\gets j+1$ because the value of the target qubit moves to $v_{i_{j+1}}$. Then, we go to Step 2.
     \item[] \textbf{Step 4.} If $v_{i_{j+2}}=v_{i_{j}}$, then we update $j\gets j+2$. Note that here we stay on the same vertex of the graph $G$. Then, we go to Step 2. 
\end{itemize}

Finally, we obtain $\textsc{ConsrtuctQFTForPath}(P',r)$ procedure whose implementation is presented in Algorithm \ref{alg:qft-path}. This procedure constructs the $r$-th part (cascade) of the circuit for QFT for the path $P'$.

\begin{algorithm}[H]
\caption{Implementation of $\textsc{ConsrtuctQFTForPath}(P',k',r)$ procedure. Algorithm of constructing the circuit for the $r$-th cascade for the path $P'=(v_{i_1}, \dots,v_{i_{k'}})$}\label{alg:qft-path}
\begin{algorithmic}
\State $j\gets 1$
\State $\textsc{H}(v_{i_j})$
\State $U\gets \emptyset$
\While{$j\leq k'-1$}
\If {$v_{i_{j+1}}\not\in U$}
\State $d\gets T_{i_{j+1}}-r$
\State  $\textsc{cP}(v_{i_{j+1}},v_{i_j},d)$
\State $U\gets U\cup\{v_{i_{j+1}}\}$
\EndIf
\If{$j\leq k'-2$ and $v_{i_{j+2}}= v_{i_{j}}$}
\State $j\gets j+2$
\Else
\If{$k'\neq 2$}
\State $\textsc{swap}(v_{i_j},v_{i_{j+1}})$
\State $w_1\gets T_{i_j}, w_2\gets T_{i_{j+1}}$
\State $Q_{w_1}\gets i_{j+1}$, $Q_{w_2}\gets i_{j}$ 
\State $T_{i_j}\gets w_2$, $T_{i_{j+1}}\gets w_1$
\EndIf
\State $j\gets j+1$
\EndIf
\EndWhile
\end{algorithmic}
\end{algorithm}

\subsection{The CNOT cost of the Circuit}\label{sec:qft3}

\begin{figure}[H]
\includegraphics[width=0.4\textwidth]{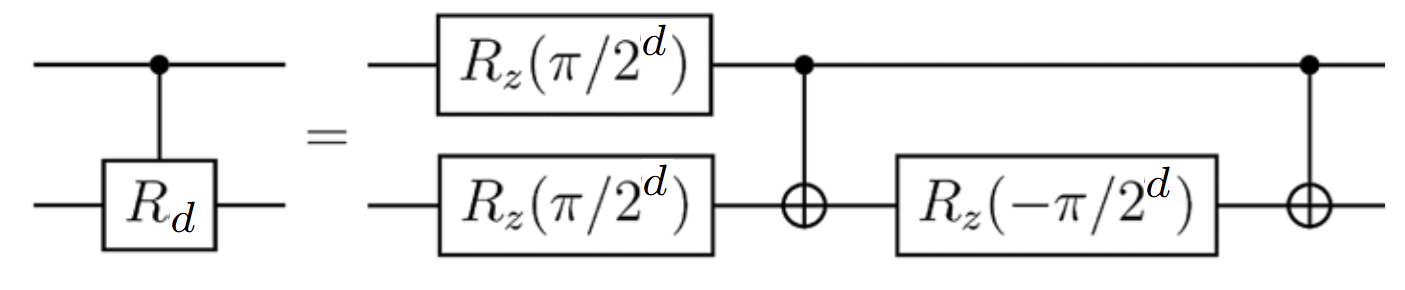}
\caption{\label{fig:cp} Representation of $CR_d$ gate using only basic gates}
\end{figure}

Note that the $CR_d$ gate can be represented using only two CNOT gates and three $R_z$ gates similar to $CR_y$ \cite{barenco1995elementary} (See Figure \ref{fig:cp}).

A pair $CR_d$ and $SWAP$ can be represented using three CNOT gates. (See Figure \ref{fig:crswap2})

\begin{figure}[H]
\includegraphics[width=0.4\textwidth]{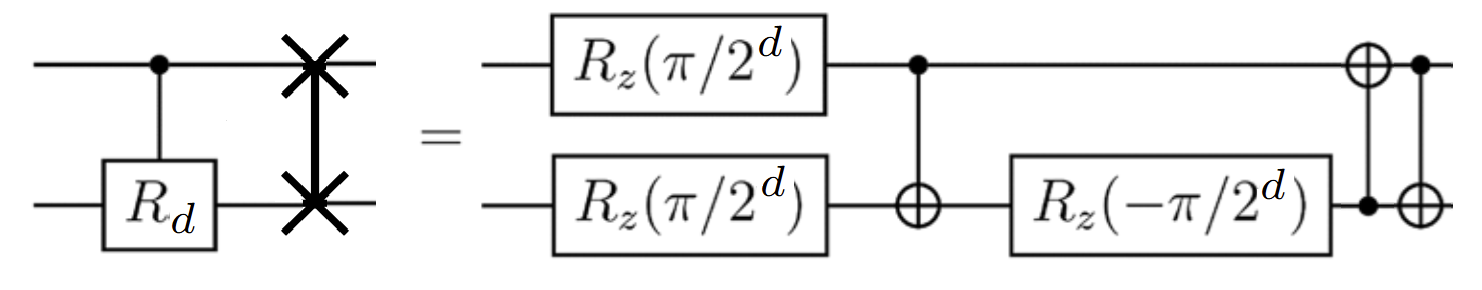}
\caption{\label{fig:crswap} Reduced representation of a pair $R_d$ and $SWAP$ gates using only basic gates}
\end{figure}  

Let us discuss the CNOT cost of the algorithm in the next theorem.
\begin{theorem}\label{th:qft}
    The CNOT cost of the circuit that is generated using Algorithm \ref{alg:qft-whole} is at most $ 3kn-2n$, where $k$ is the length of the shortest non-simple path $P=(v_{i_1},\dots,v_{i_k})$ in the graph $G$ that visits all vertices at least once.
\end{theorem}
\Beginproof
We can say that each of the paths for each cascade has a length at most $k$, where $P=(v_{i_1},\dots,v_{i_k})$ is the shortest path that visits all vertices at least once. 

Similarly to the proof of Theorem \ref{th:qh1}, we can show that the CNOT cost of one cascade is at most $3k-2$.  So, the total number of CNOT gates is at most $3nk-2n$.
\Endproof

\begin{theorem}\label{th:qft-hamiltonian}
If the graph $G$ with $n$ vertices has a Hamiltonian path, then the CNOT cost of the produced circuit for the QFT algorithm is $1.5n^2-1.5n-1$.
\end{theorem}
\Beginproof
Because the graph has a Hamiltonian path, the path $P=(v_{i_1},\dots,v_{i_n})$ is exactly this path. So, on each step, we move the target qubit. We can say that in constructing the $r$-th cascade, the path $P'=(v_{i_1},\dots,v_{i_{n-r+1}})$.
The CNOT cost of the $r$-th cascade for $1\leq r\leq n-2$ is $3(n-r)$ because we apply SWAP and the control phase gates for each edge of the path. The CNOT cost of the $(n-1)$-th cascade is $2$ because we do not apply the SWAP gate. The CNOT cost of the $n$-th cascade is $0$. Therefore, the total cost is $\sum_{r=1}^{n-1}(3(n-r))-1=1.5n^2-1.5n-1$.
\Endproof

Using Lemma \ref{lm:path-len} we can estimate the CNOT cost in the next corollary.

\begin{corollary}\label{cr:path2}
The CNOT cost of a circuit that is generated using Algorithm \ref{alg:qh} is in the range from $1.5n^2-1.5n-1$ to $1.5n^3-1.5n^2-2n$. 
\end{corollary}
\Beginproof
If $k=n$, then the CNOT cost is at most
$1.5n^2-1.5n-1$ because of Theorem \ref{th:qft-hamiltonian}.

If $k=n(n-1)/2$ that is the maximum possible value for $k$, then the
CNOT cost is at most 
\[1.5n^2(n-1)-2n=1.5n^3-1.5n^2-2n.\]
\Endproof

\subsection{Comparing With Other Results}\label{sec:qft4}
Similarly to Section \ref{sec:compare-qh} we discuss the LNN architecture and more complex architecture.
Firstly, let us consider the LNN architecture. As was discussed earlier (see Figure \ref{fig:lnn-5}), the path visits all vertices from $v_1$ to $v_n$ one by one. The paths constructed for each cascade are presented in Appendix \ref{apx:lnn-qft}. The circuit is similar to the circuit developed in \cite{k2024aliya}. The presented path is the Hamiltonian path in the graph. Due to Theorem \ref{th:qft-hamiltonian}, we get the following  CNOT cost for the LNN architecture.
\begin{lemma}
The CNOT cost of the produced circuit of the QFT algorithm using $n$ qubits for the LNN architecture is $1.5n^2-1.5n-1$.
\end{lemma}


At the same time, \cite{park2023reducing} gives the circuit with $n^2+n-4$ CNOT cost, and \cite{k2024aliya} gives the circuit with $1.5n^2-2.5n+1$ CNOT cost. Our circuit is better than \cite{park2023reducing} only if $n\leq 3$. The result from \cite{k2024aliya} is better than our result. 
However, it is not known how to apply the results of \cite{park2023reducing} for more complex architecture, and the results of \cite{k2024aliya} are presented only for a specific architecture.

Secondly, let us consider more complex architectures like 16-qubit ``sun''  (Figure \ref{fig:sun1}), and 27-qubit  ``two joint suns'' (Figure \ref{fig:cairo1}). The paths are presented in figures. Note that the red part of the path is such that $v_{j+2}=v_{j}$. Therefore, we do not apply a SWAP gate on this step, but only the control phase operator. The CNOT cost for 16-qubit machine is $342$, and for 27-qubit machine is $1009$

So, our generic method gives a bit worse circuits than the circuits constructed for these devices especially \cite{k2024aliya}, which CNOT costs are $324$ and $957$ for 16-qubit and 27-qubit architectures, respectively. The difference is about 5\%.
\section{Conclusion}\label{sec:concl}
We present a generic method for constructing quantum circuits for quantum fingerprinting (quantum hashing) and quantum Fourier transform algorithms for an arbitrary connected qubit connection graph for hardware. The heuristic version of the method is fast enough and works in $O(n^5)$ time complexity. The certain version has an exponential time complexity that is $O(n^22^n)$. The algorithm works for arbitrary connection graph.

At the same time, if we consider samples of graphs like LNN, ``sun'' (16-qubit IBMQ Eagle r3 architecture), ``two joint suns'' (27-qubit IBMQ Eagle r3 architecture), then our generic algorithm gives us the same circuit as the techniques provided especially for these graphs \cite{k2024aliya,ksy2024} in the case of quantum hashing. In the case of QFT, our algorithm gives a bit worse circuit compared to the techniques optimized for these graphs \cite{park2023reducing,k2024aliya}. At the same time, our approach works for arbitrary connected graphs, but the existing results work only for some specific graphs.

An open question is to develop a technique for QFT for an arbitrary connected graph that gives us the same or better results than the existing ones for specific architectures like LNN, ``sun'' and ``two joint suns''.
\begin{acknowledgments}
This work is supported by the National Natural Science Foundation of China under Grant No.
61877054, 12031004,12271474, and the Foreign Experts in Culture and Education Foundation
under Grant No. DL2022147005L. Kamil Khadiev thanks these projects for supporting his
visit. The research by Kamil Khadiev in Section \ref{sec:qft} is supported by Russian Science Foundation Grant 24-21-00406, https://rscf.ru/en/project/24-21-00406/.
The study was funded by the subsidy allocated to Kazan Federal University for the state assignment in the sphere of scientific activities (Project No. FZSM-2024-0013).
\end{acknowledgments}

\appendix
\section{Implementation of the Procedure $\textsc{ShortestPathes}$ for Shortest Paths Searching}\label{apx:floyd}
Here we discuss how to construct matrices $W$ and $A$ such that $W[v,u]$ is the length of the shortest path between vertices $v$ and $u$, and $A[v,u]$ is the last vertex in the shortest path between $v$ and $u$. The procedures are simple, but we present them for the completeness of the results representation.

Firstly, we present a procedure $\textsc{SingleSrcShortestPath}(v)$ that finds the shortest paths for a single source vertex $v$ that is based on the BFS algorithm \cite{cormen2001}. The algorithm calculates the $v$-th rows of $W$ and $A$. The implementation is presented in Algorithm \ref{alg:bfs1}. Here we assume that we have a queue data structure \cite{cormen2001} that allows us to do the next actions in constant time:
\begin{itemize}
    \item $\textsc{Add}(queue, v)$ adds an element to the queue;
    \item $\textsc{Remove}(queue)$ removes an element from the queue and returns the element;
    \item  $\textsc{Init}()$ returns an empty queue;
     \item $\textsc{isEmpty}(queue)$ returns $True$ if the queue is empty and $False$ otherwise.
\end{itemize}
\begin{algorithm}[H]
\caption{Implementation of $\textsc{SingleSrcShortestPath}(v)$}\label{alg:bfs1}
\begin{algorithmic}
\State $queue\gets \textsc{Init}()$
\State  $\textsc{Add}(queue, v)$
\For{$u\in V$}
\State $W[v,u]\gets\infty$
\State $A[v,u]\gets NULL$
\EndFor
\State $W[v,v]\gets 0$
\While{$\textsc{isEmpty}(queue)=False$}
\State $t\gets\textsc{Remove}(queue)$
\For{$r\in \textsc{Neighbors}(t)$}
\If{$W[v,r]=\infty$}
    \State $A[v,r]\gets t$
    \State $W[v,r]=W[v,t]+1$
    \State $\textsc{Add}(queue, r)$
\EndIf
\EndFor
\EndWhile
\end{algorithmic}
\end{algorithm}
As an implementation of the $\textsc{ShortestPaths}$ procedure, we invoke $\textsc{SingleSrcShortestPath}(v)$ for each vertex $v\in V$.
\begin{algorithm}[H]
\caption{Implementation of $\textsc{ShortestPaths}(G)$ for a $G=(V,E)$ graph}\label{alg:bfs2}
\begin{algorithmic}
\State 
\For{$v\in V$}
    \State $\textsc{SingleSrcShortestPath}(v)$
\EndFor
\State \Return $(W,A)$
\end{algorithmic}
\end{algorithm}
\begin{lemma}
    Time complexity of the $\textsc{ShortestPathes}$ procedure is $O(n^3)$.
\end{lemma} 
\Beginproof
Time complexity of BFS is $O(n+m)=O(n^2)$ due to \cite{cormen2001}.  Invocation of $n$ BFS algorithms for each $v\in V$ is $O(n^3)$. 
\Endproof
\section{Quantum Fingerprinting or Quantum Hashing}\label{apx:hash}
Let us present some basic concepts of quantum fingerprinting technique from \cite{af98, an2008,an2009,akv2008,bcwd2001}. This technique is well-known and allows us to compute a short hash or fingerprint that identifies an original large data with high probability.

For the problem being solved we choose some positive integer $m$, an error probability bound $\varepsilon > 0$, fix $t = \lceil(2/\varepsilon) \ln 2m\rceil$, and construct a mapping $g : \{0, 1\}^n\to \mathbb{Z}$. Then for arbitrary binary string $\sigma = (\sigma_1 \dots \sigma_n)$ we create it's fingerprint $|h_\sigma\rangle$ composing $t$ single qubit fingerprints $|h_\sigma^i\rangle$:
\[|h_\sigma^i\rangle=cos\frac{2\pi k_i g(\sigma)}{m}|0\rangle + sin\frac{2\pi k_i g(\sigma)}{m}|0\rangle,\]\[
|h_\sigma\rangle=\frac{1}{\sqrt{t}}\sum_{i=1}^{t}|i\rangle|h^i_{\sigma}\rangle\]

Here the last qubit is rotated by $t$ different angles about the $\hat{y}$ axis of the Bloch sphere. The chosen parameters $k_i \in\{1\dots,m-1\}$, for $i\in\{1\dots t\}$ are ``good'' in the following sense. A set of parameters $K = \{k_1,\dots, k_t\}$ is called ``good'' for $g\neq 0 \mod m$ if
\[\frac{1}{t^2}\left(\sum_{i=1}^t cos\frac{2\pi k_i g}{m}\right)^2<\varepsilon\]
The left side of the inequality is the squared amplitude of the basis state $|0\rangle^{\otimes \log_2 t} |0\rangle$ if the operator
$H^{\otimes \log_2 t}\otimes I $ has been applied to the fingerprint $|h_\sigma\rangle$. Informally, that kind of set guarantees, that
the probability of error will be bounded by a constant below $1$.

The following lemma from \cite{akv2008,an2008,an2009} proves the existence of a ``good'' set.
\begin{lemma}[\cite{akv2008}]
There is a set $K$ with $|K| = t = \lceil(2/\varepsilon) \ln 2m\rceil$  which is ``good'' for all $g\neq 0 \mod m$.
\end{lemma}

We use this result for fingerprinting technique \cite{akv2008} choosing the set $K = \{k_1,\dots, k_t\}$ that is ``good'' for all $g = g(\sigma)\neq 0$. It allows us to distinguish those inputs whose image is $0$ modulo $m$ from the others.

That hints at how this technique may be applied:
\begin{enumerate}
\item We construct $g(x)$, that maps all acceptable inputs to $0$ modulo $m$ and others to arbitrary non-zero (modulo $m$) integers.

\item After the necessary manipulations with the fingerprint, the $H^{\otimes \log_2 t}$ operator is applied to the first $\log_2 t$ qubits. This operation ``collects'' all cosine amplitudes at the all-zero state. That is, we obtain the state of the type
\[|h'_\sigma\rangle=\frac{1}{t}\sum_{i=1}^{t}cos\left(\frac{2\pi k_i g(\sigma)}{m}\right) |00\dots 0\rangle|0\rangle + \sum_{i=2}^{2t}\alpha_i|i\rangle\]
\item This state is measured on the standard computational basis. Then we accept the input if the outcome is the all-zero state. This happens with the probability
\[Pr_{accept}(\sigma)=\frac{1}{t^2} \left(\sum_{i=1}^{t}cos\frac{2\pi k_i g(\sigma)}{m}\right)^2,\]
which is $1$ for the inputs, whose image is $0 \mod m$ and is bounded by $\varepsilon$ for the others.
\end{enumerate}

Due to \cite{ziiatdinov2023gaps,kalis18}, the algorithm can be implemented using the shallow circuit presented in Figure \ref{fig:qf}. In that case, the angles $\frac{2\pi k_i}{m}$ should be linear combinations of $\xi_j$. Due to \cite{ziiatdinov2023gaps}, it is not known whether we can keep the same number of qubits or should we increase the number of qubits exponentially. At the same time, computational experiments show \cite{ziiatdinov2023gaps} that we can find enough good parameters $\xi_i$ such that $t$ qubits are enough for $\varepsilon$ error probability.

\section{Quantum Fourier Transform}\label{apx:qft}
QFT is a quantum version of the discrete Fourier transform. The definition of $n$-qubit QFT and its inverse are as follows:
\[QFT|j\rangle = \sum_{k=0}^{2^n-1}e^{\frac{2\pi i jk}{2^n}}|k\rangle,\]
\[QFT^{-1}|j\rangle = \sum_{k=0}^{2^n-1}e^{-\frac{-2\pi i jk}{2^n}}|k\rangle,\]
The $n$-qubit QFT circuit requires $0.5n^2 - 0.5n$ control phase ($CR_d$) gates and $n$ Hadamard ($H$) gates if we have no restriction on the application of two-qubit gates (See Figure \ref{fig:cqft}). The $CR_d$ gate is represented by basic gates that require two CNOT and three $R_z$ gates \cite{barenco1995elementary}. Therefore, $n^2 - n$ CNOT gates are required to construct an $n$-qubit QFT circuit. At the same time, if a quantum device has the LNN architecture, then for implementing the QFT, the number of CNOT gates is much larger than $n^2 - n$ \cite{fdh2004,swd2011,wld2014,kds2017,bbwdr2019,park2023reducing}. If we consider a general graph, then the situation is much worse \cite{k2024aliya}. 

\section{Constructing the List of First Indexes}\label{apx:firstindexes}
For a path $P=(v_{i_1},\dots,v_{i_k})$ we construct a list $(j_1,\dots, j_k)$ such that $j_r\not\in\{i_1,\dots i_{j_r-1}\}$.
We use a set $U'$ that is implemented using the Red-Black tree \cite{cormen2001}.  Let $J$ be the results list.

Initially, $U'\gets\emptyset$ is an empty set, and $J\gets()$ is an empty list.
For each $h$ from $1$ to $k$ we check whether $i_h\in U'$. If $i_h\not\in U'$, then we add $i_h$ to $J$, otherwise, we skip it.

\begin{algorithm}[H]
\caption{Constructing the list $J=(j_1,\dots,j_n)$ by the path $P$}
\begin{algorithmic}
\State $U'\gets \emptyset$
\State  $J\gets ()$
\For{$h\in \{1,\dots,k\}$}
\If{ $i_h\not\in U'$}
\State $U'\gets U'\cup\{i_h\}$
\State $J\gets J\circ i_h$
\EndIf
\EndFor
\State \Return $J$
\end{algorithmic}
\end{algorithm}

\section{Graphs for Each Cascade for LNN Architecture}
\label{apx:lnn-qft}
\begin{figure}[H]
\includegraphics[width=0.3\textwidth]{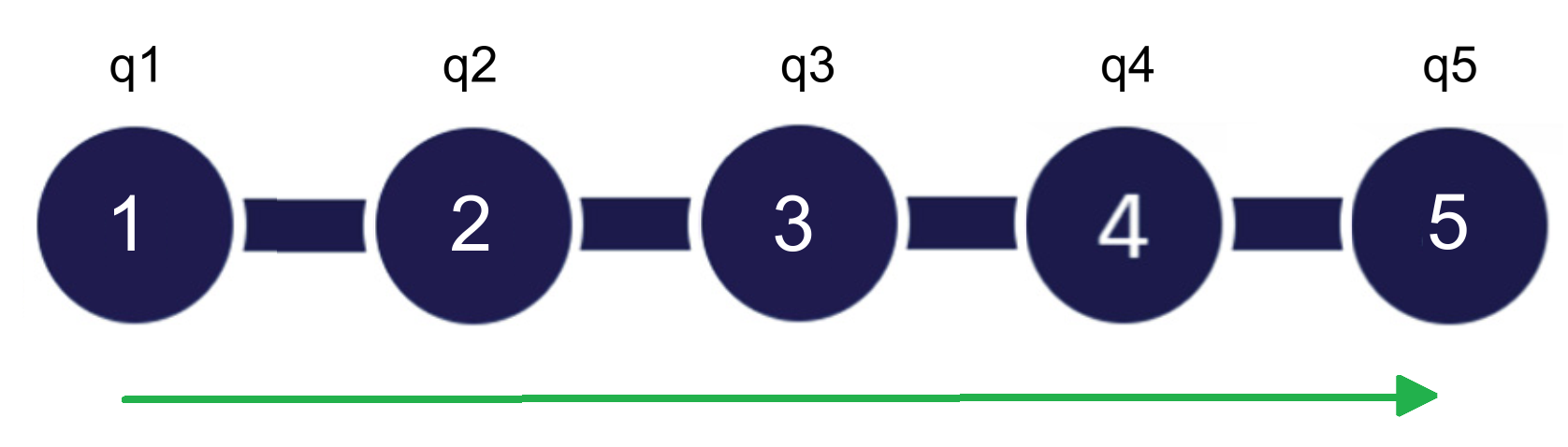}
\caption{ The graph for the 5-qubit LNN architecture. The path that visits all vertices at least once is green. $q1,\dots,q5$ are assigned logical qubits for the vertices.}
\label{fig:lnn5-1}
\end{figure}

\begin{figure}[H]
\includegraphics[width=0.45\textwidth]{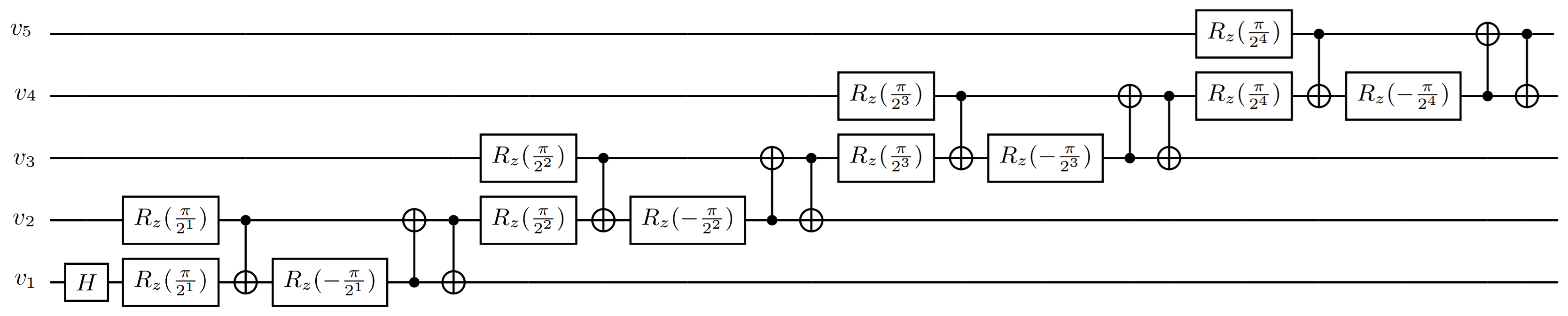}
\caption{The circuit for the first cascade.}
\label{fig:lnn5-1-c}
\end{figure}

The path for the the 5-qubit LNN architecture is presented in Figure \ref{fig:lnn5-1}. It is used for the first cascade, the circuit of which is presented in Figure \ref{fig:lnn5-1-c}.

\begin{figure}[H]
\includegraphics[width=0.3\textwidth]{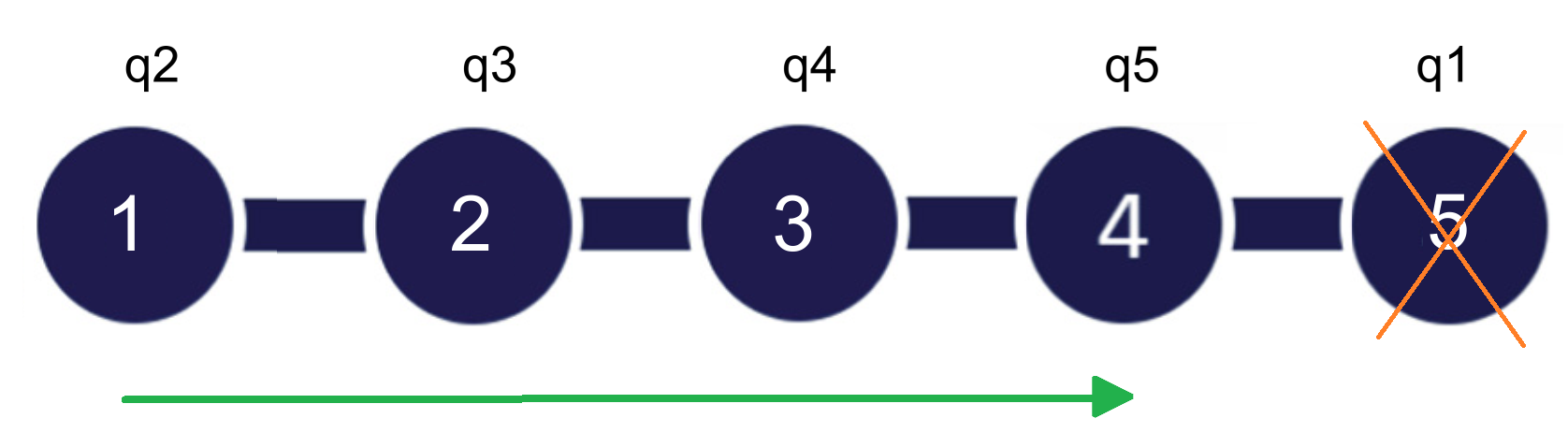}
\caption{ The graph for the 5-qubit LNN architecture. The second cascade excludes the vertex that corresponds to $1$-st logical qubit. The path that visits all vertices at least once is green.}
\label{fig:lnn5-2}
\end{figure}

\begin{figure}[H]
\includegraphics[width=0.45\textwidth]{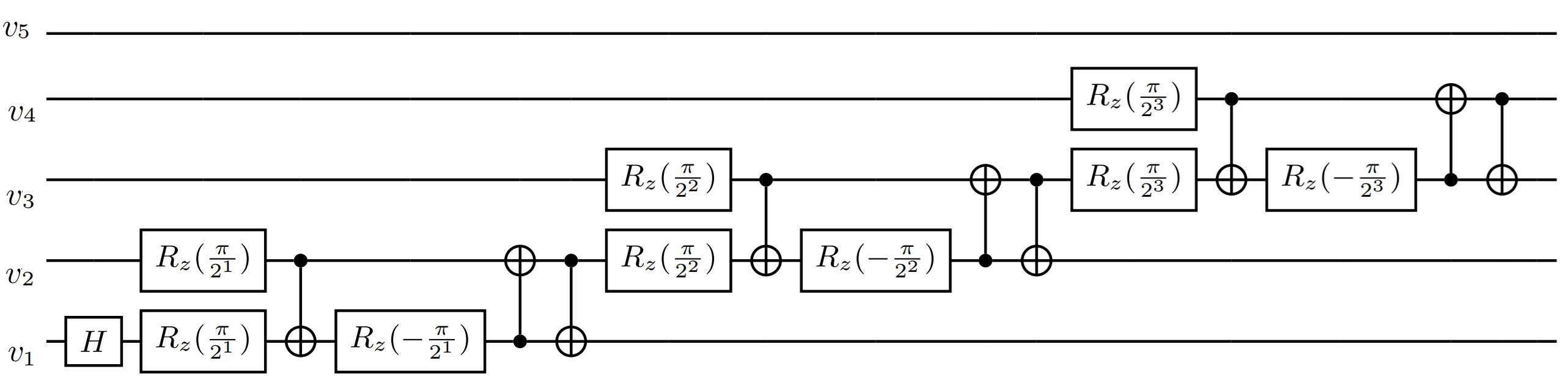}
\caption{The circuit for the second cascade.}
\label{fig:lnn5-2-c}
\end{figure}

The path for the 2-d cascade is presented in Figure \ref{fig:lnn5-2} and the corresponding circuit is in Figure \ref{fig:lnn5-2-c}. 

\begin{figure}[H]
\includegraphics[width=0.3\textwidth]{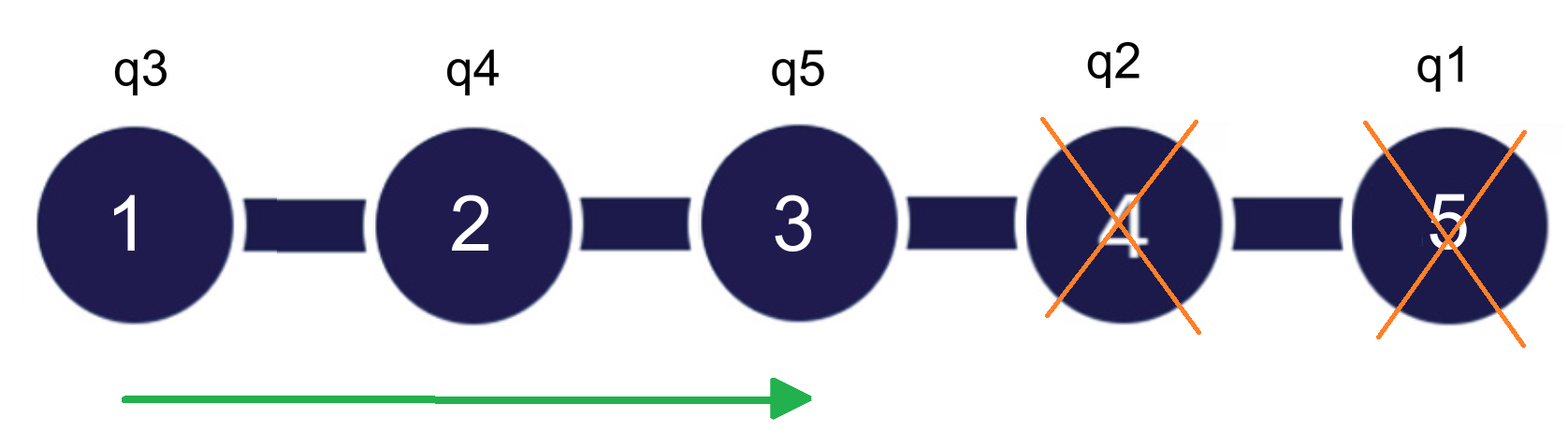}
\caption{ The graph for the 5-qubit LNN architecture. The $3$-d cascade excludes the vertices that correspond to $1$-st and $2$-nd  logical qubits. The path that visits all vertices at least once is green.}
\label{fig:lnn5-3}
\end{figure}

\begin{figure}[H]
\includegraphics[width=0.3\textwidth]{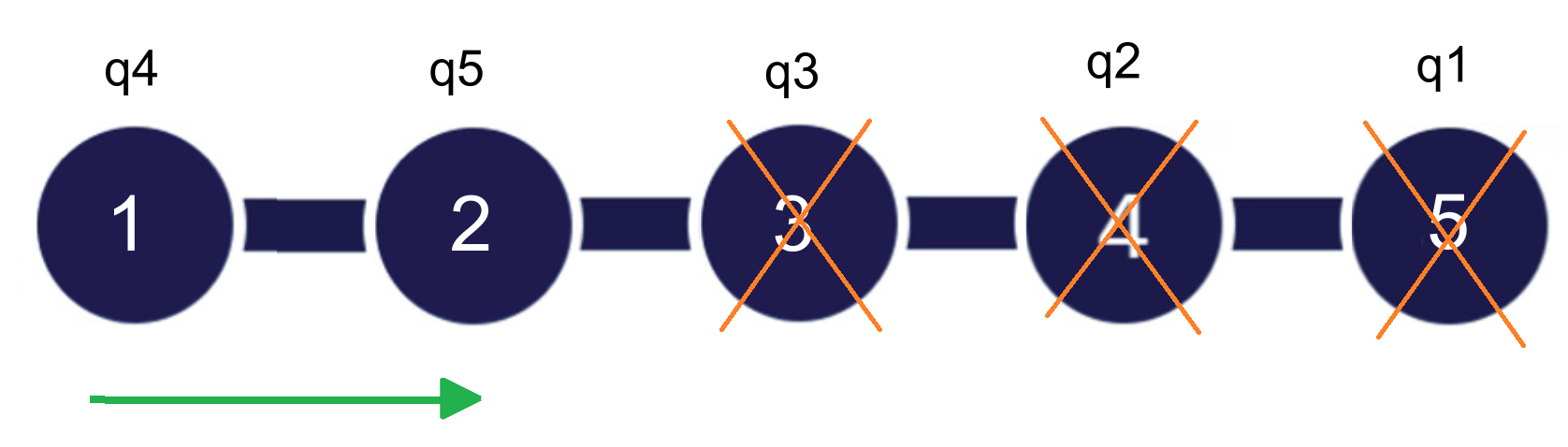}
\caption{ The graph for the 5-qubit LNN architecture. The $4$-th cascade excludes the vertices that correspond to $1$-st, $2$-nd, and $3$-d  logical qubits. The path that visits all vertices at least once is green.}
\label{fig:lnn5-4}
\end{figure}

\begin{figure}[H]
\includegraphics[width=0.45\textwidth]{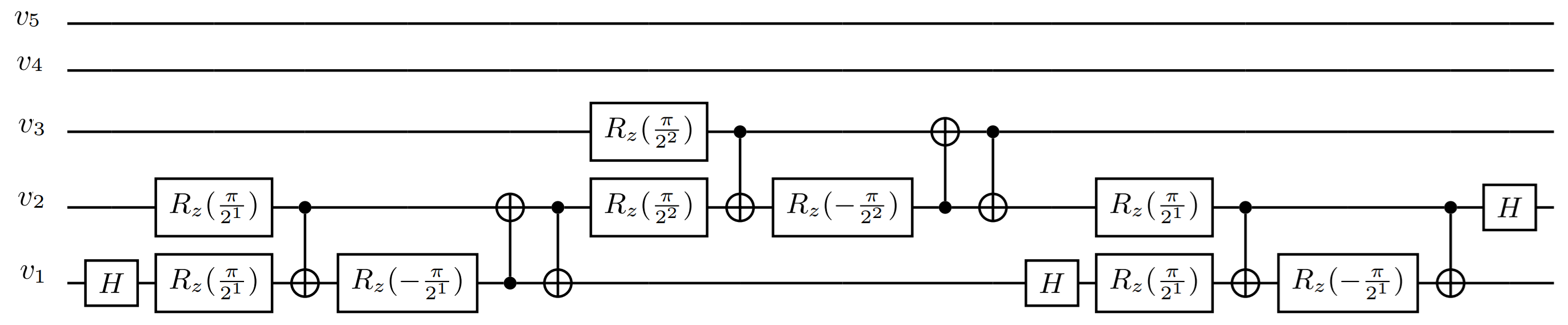}
\caption{The circuit for the 3-d, 4-th, and 5-th cascades.}
\label{fig:lnn5-3-c}
\end{figure}

The paths for the 3-d and 4-th cascades are presented in Figures \ref{fig:lnn5-3} and \ref{fig:lnn5-4}. The corresponding circuit including the last 5-th cascade is in Figure \ref{fig:lnn5-3-c}.

\section{Graphs for Each Cascade for the 16-qubit ``sun'' Architecture}

The paths for each of the first 15 cascades are in the following Figures.

\begin{figure}[H]
\includegraphics[width=0.28\textwidth]{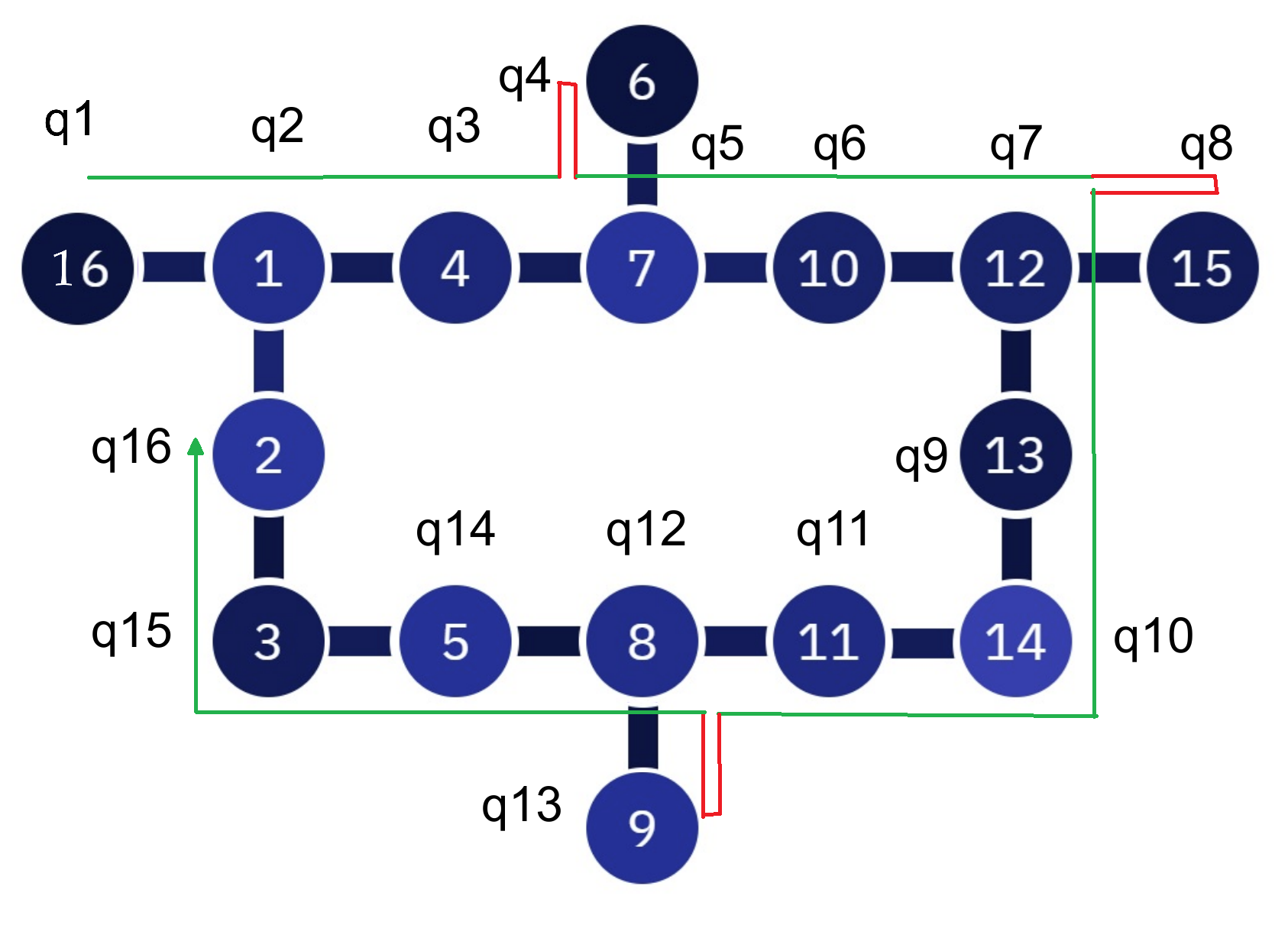}
\caption{
The graph for the 16-qubit ``sun'' architecture. The path that visits all vertices at least once is green. Red parts are such that $v_{i_j}=v_{i_{j+2}}$. The labels $q1,\dots,q16$ are names of assigned logical qubits for the vertices.}
\end{figure}

\begin{figure}[H]
\includegraphics[width=0.28\textwidth]{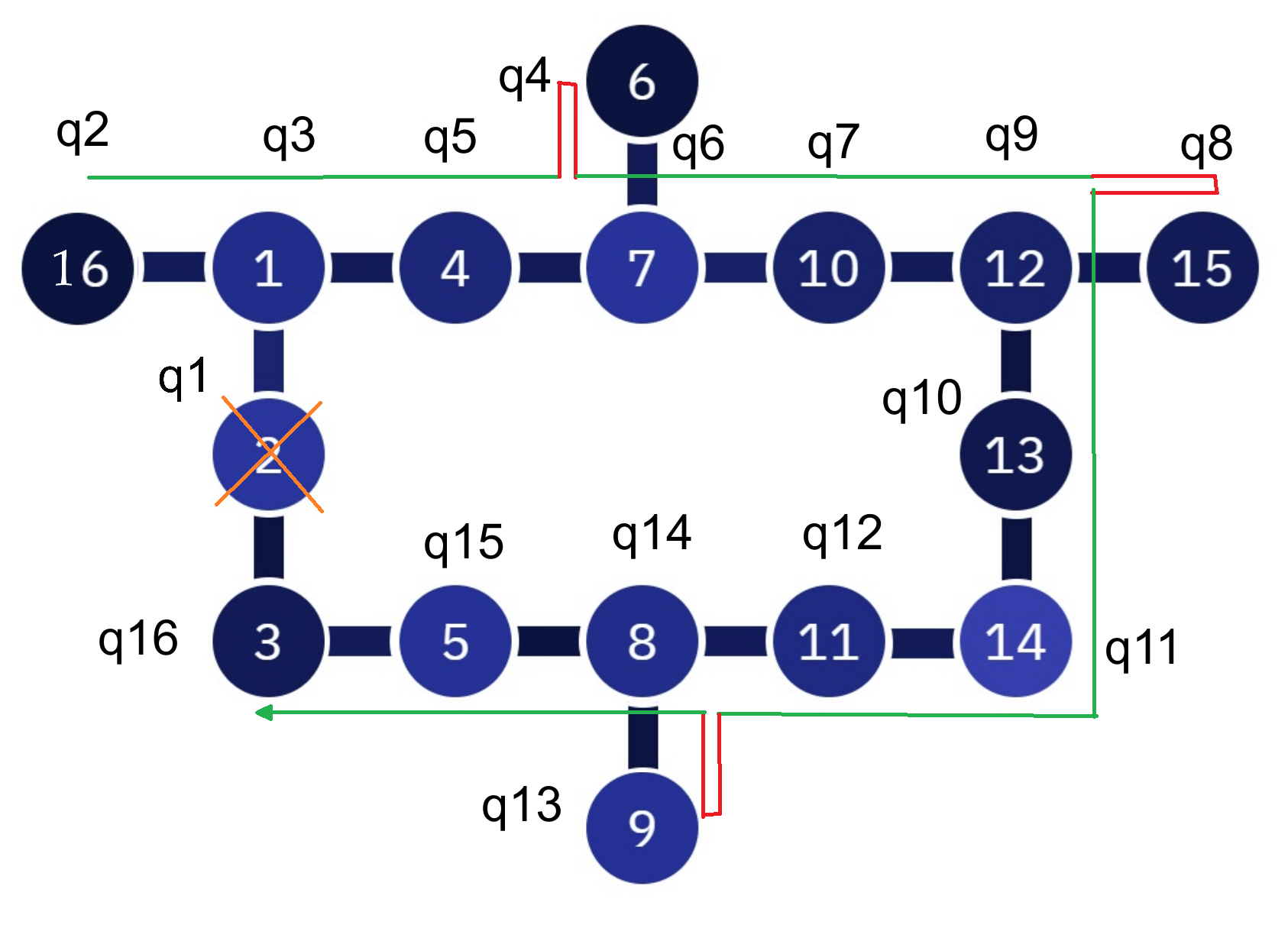}
\caption{
The graph for the 16-qubit ``sun'' architecture. The second cascade excludes the vertex that corresponds to the $1$-st logical qubit. The path that visits all vertices at least once is green. Red parts are such that $v_{i_j}=v_{i_{j+2}}$.}
\end{figure}

\begin{figure}[H]
\includegraphics[width=0.28\textwidth]{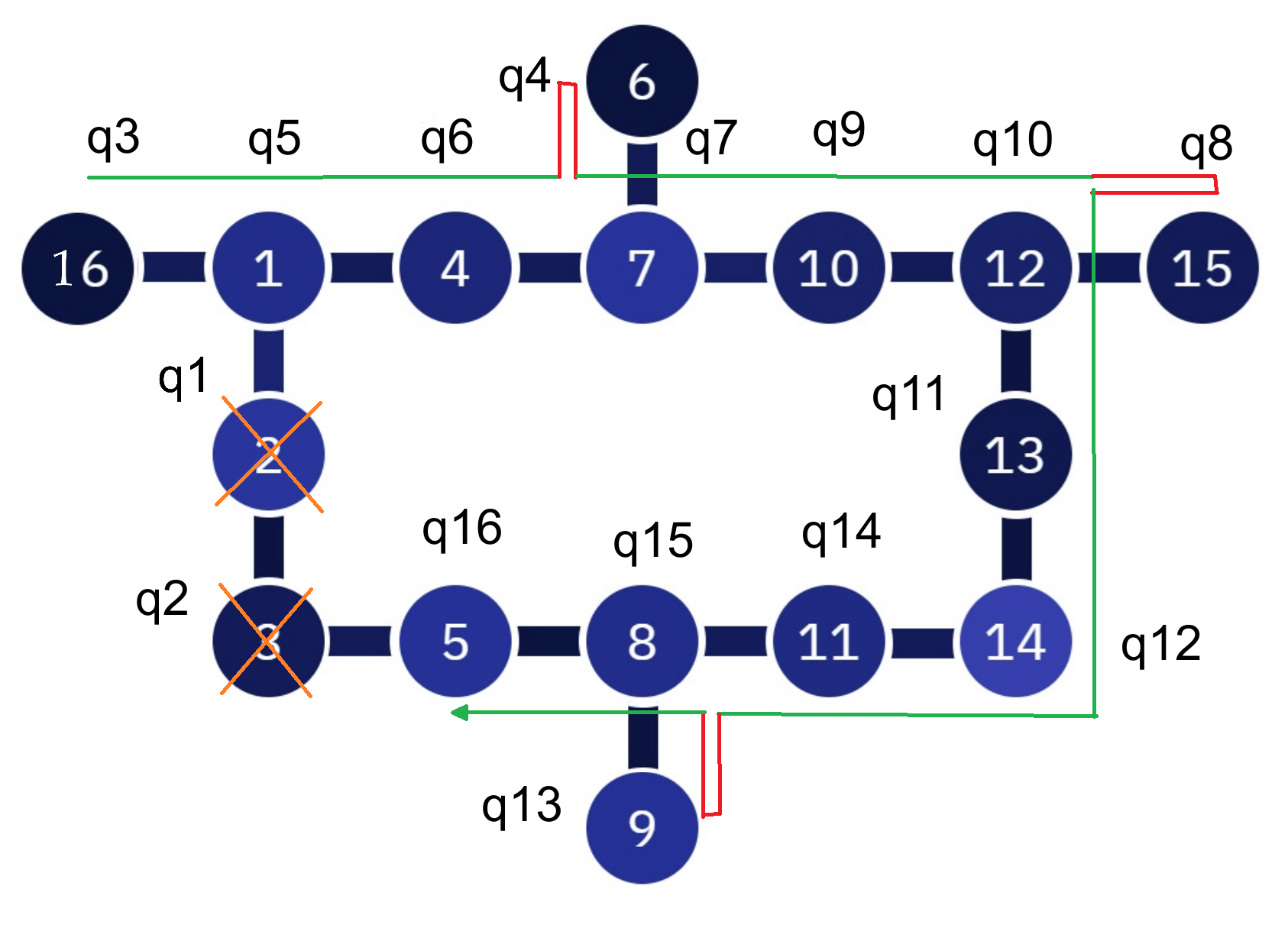}
\caption{
The graph for the 16-qubit ``sun'' architecture. The $3$-d cascade excludes the vertex that corresponds to the logical qubits with indexes from the $\{1,2\}$ set.}
\end{figure}

\begin{figure}[H]
\includegraphics[width=0.28\textwidth]{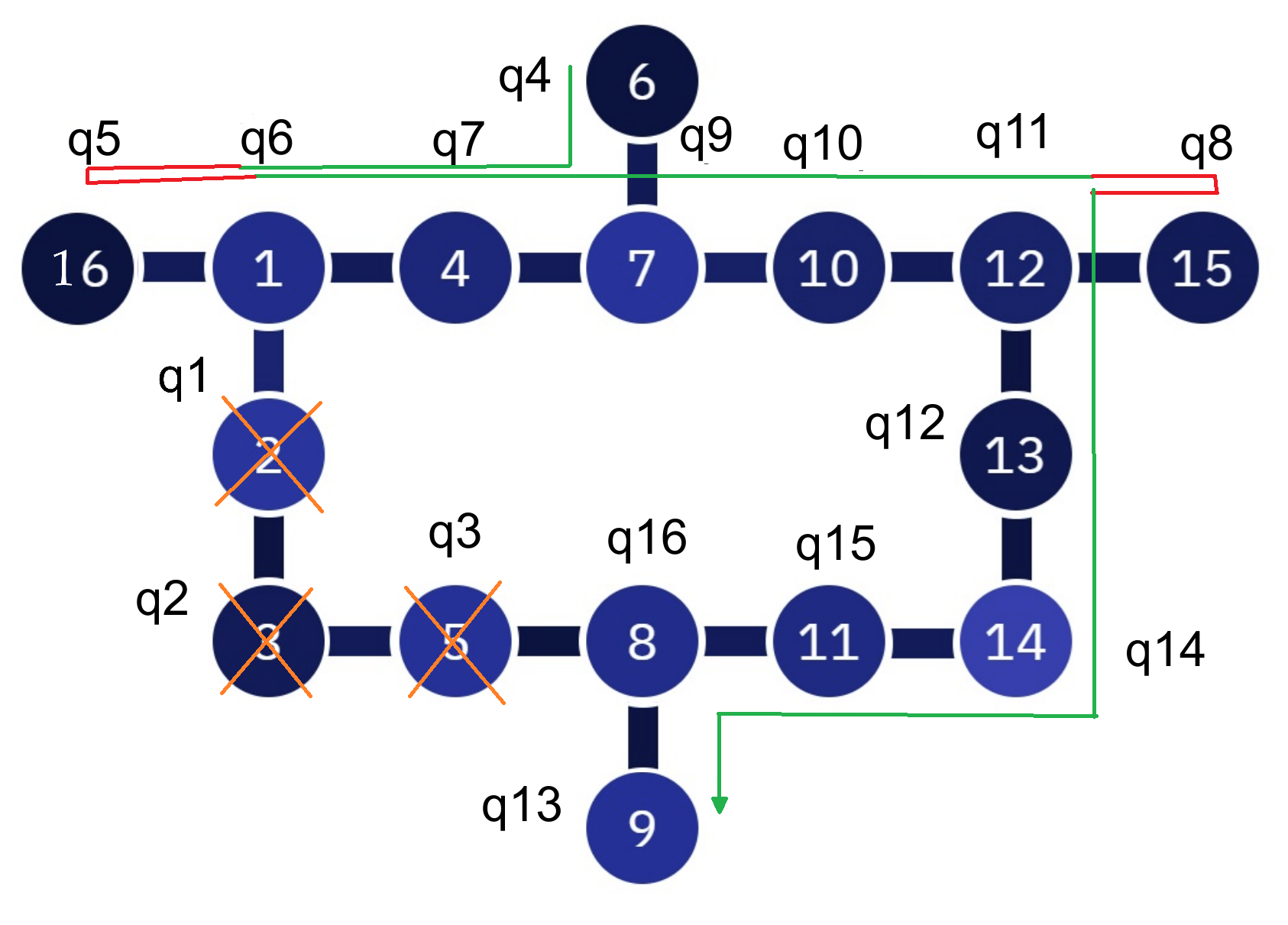}
\caption{
The graph for the 16-qubit ``sun'' architecture. The $4$-th cascade excludes the vertex that corresponds to the logical qubits with indexes from the $\{1,2,3\}$ set.}
\end{figure}

\begin{figure}[H]
\includegraphics[width=0.28\textwidth]{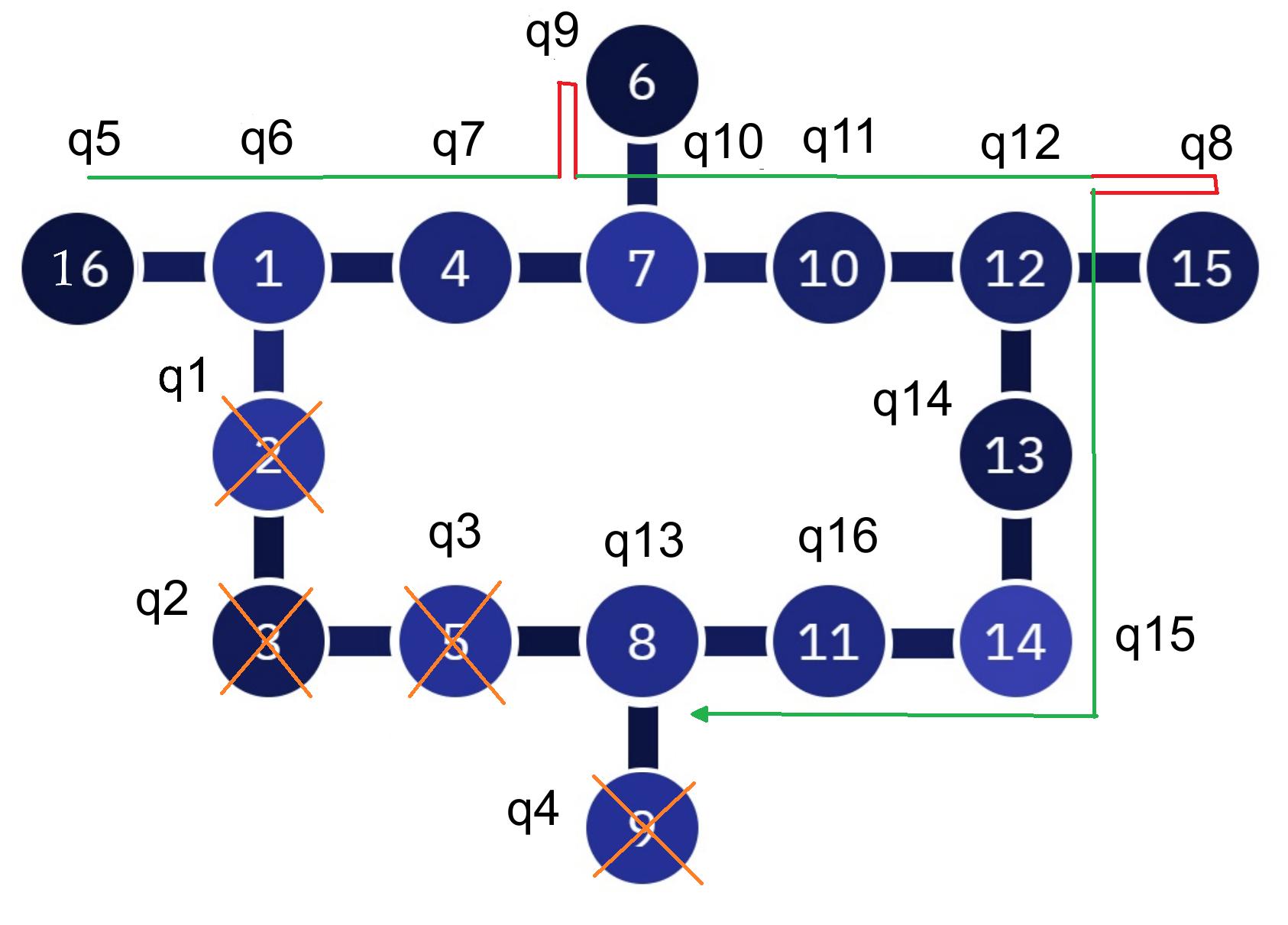}
\caption{
The graph for the 16-qubit ``sun'' architecture. The $5$-th cascade that excludes the vertex that corresponds to the logical qubits with indexes from the $\{1,\dots,4\}$ set.}
\end{figure}

\begin{figure}[H]
\includegraphics[width=0.28\textwidth]{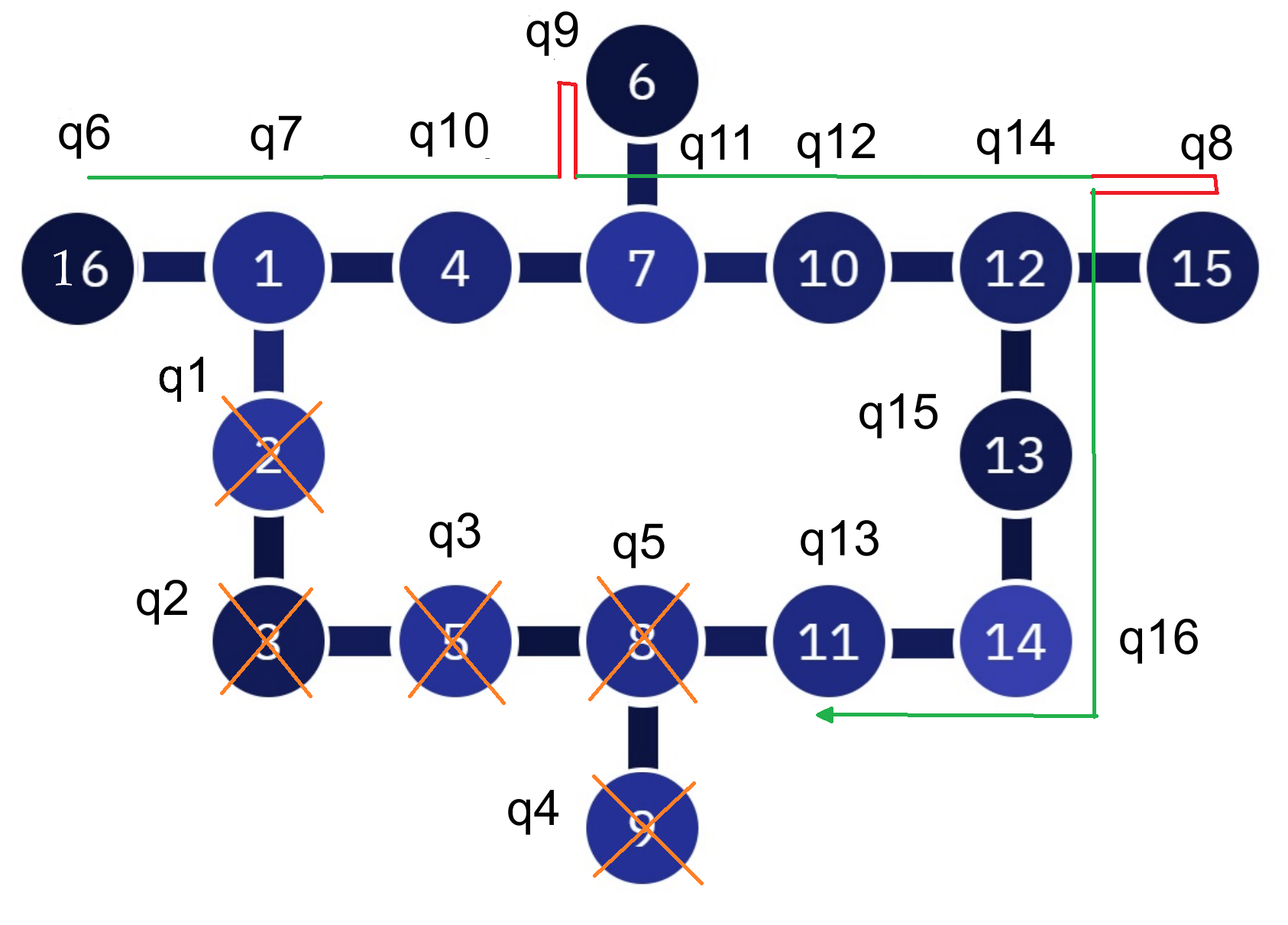}
\caption{
The graph for the 16-qubit ``sun'' architecture. The $6$-th cascade that excludes the vertex that corresponds to the logical qubits with indexes from the $\{1,\dots,5\}$ set.}
\end{figure}

\begin{figure}[H]
\includegraphics[width=0.28\textwidth]{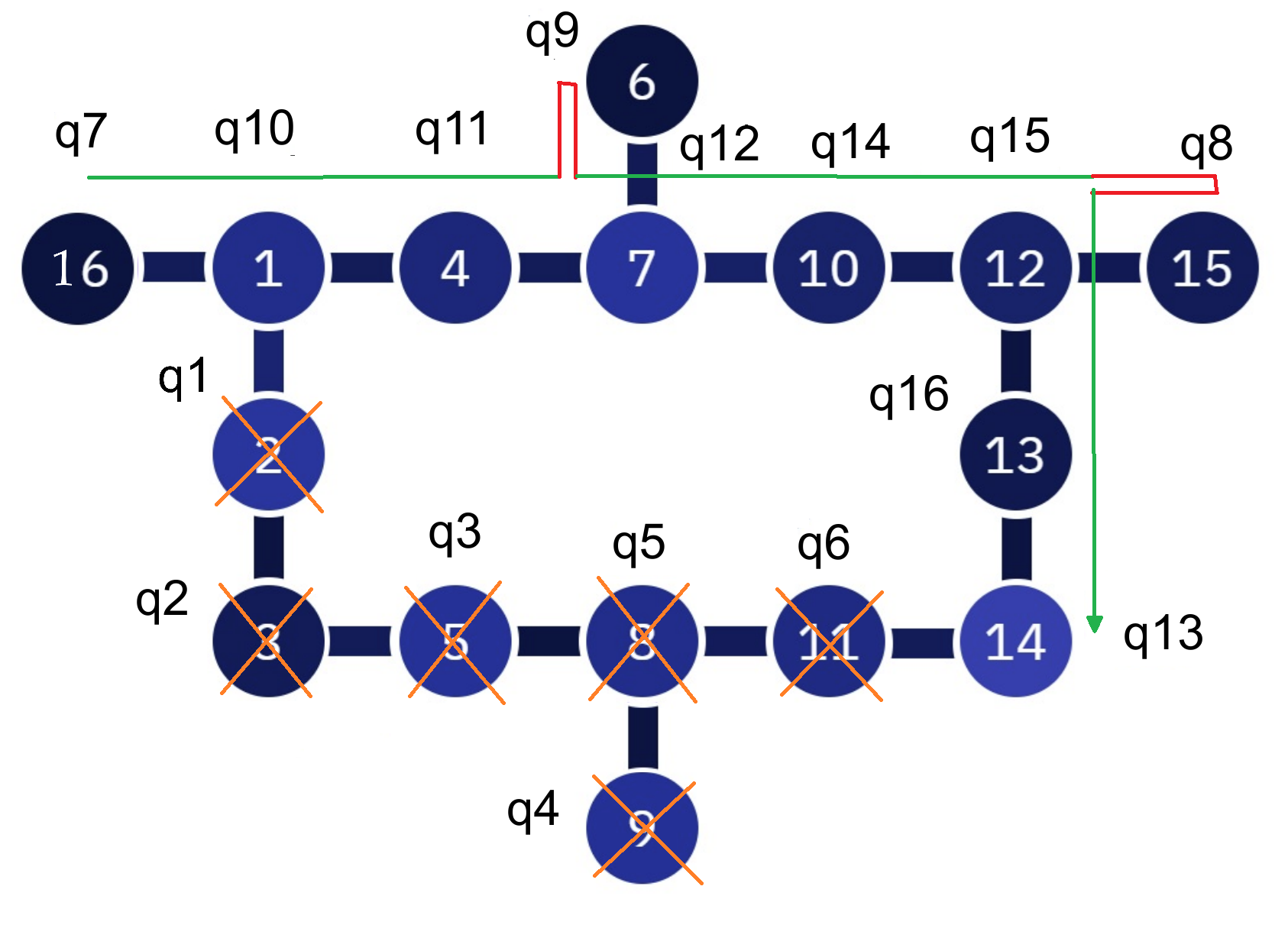}
\caption{
The graph for the 16-qubit ``sun'' architecture. The $7$-th cascade that excludes the vertex that corresponds to the logical qubits with indexes from the $\{1,\dots,6\}$ set.}
\end{figure}

\begin{figure}[H]
\includegraphics[width=0.28\textwidth]{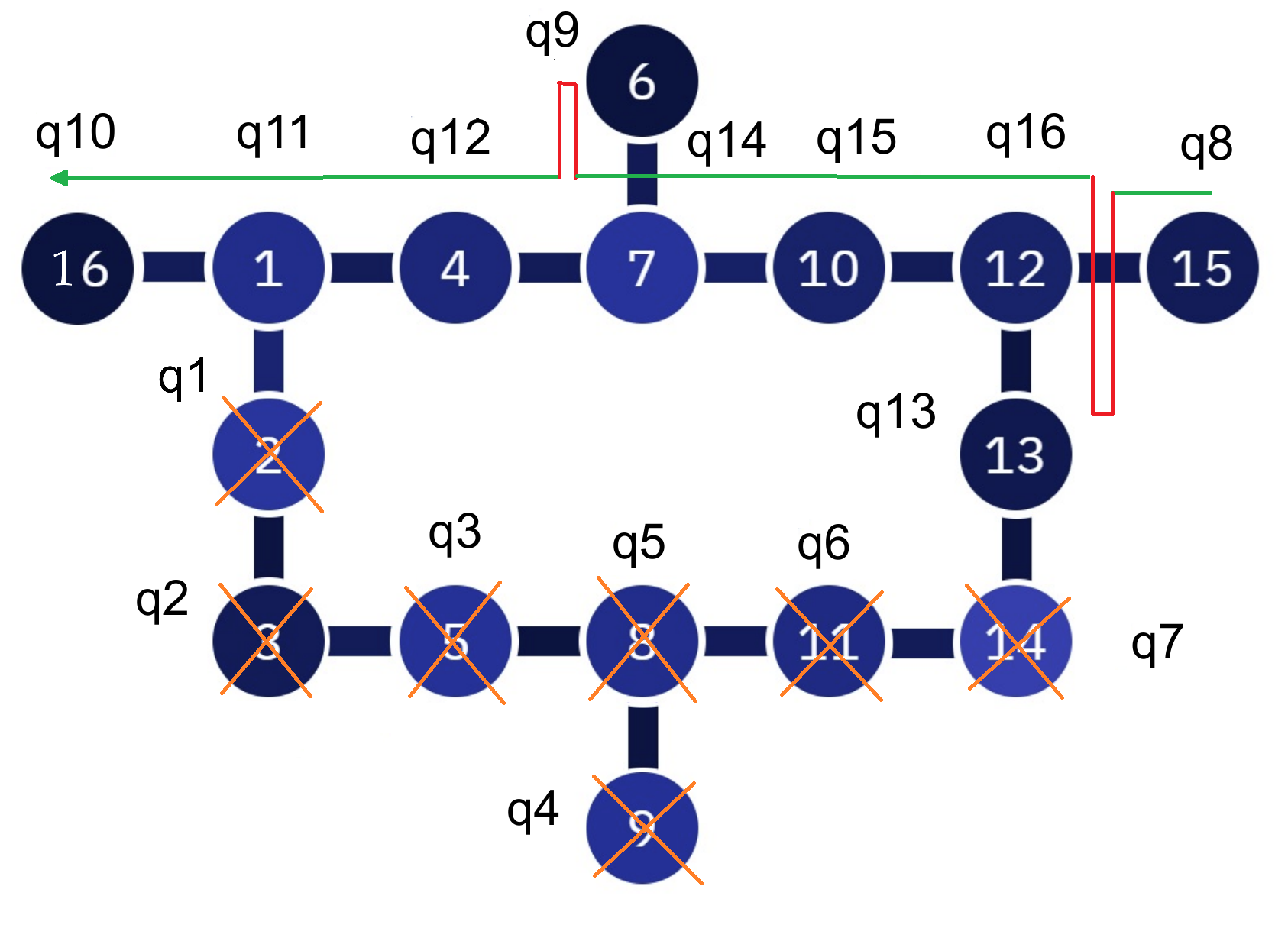}
\caption{
The graph for the 16-qubit ``sun'' architecture. The $8$-th cascade that excludes the vertex that corresponds to the logical qubits with indexes from the $\{1,\dots,7\}$ set.}
\end{figure}

\begin{figure}[H]
\includegraphics[width=0.28\textwidth]{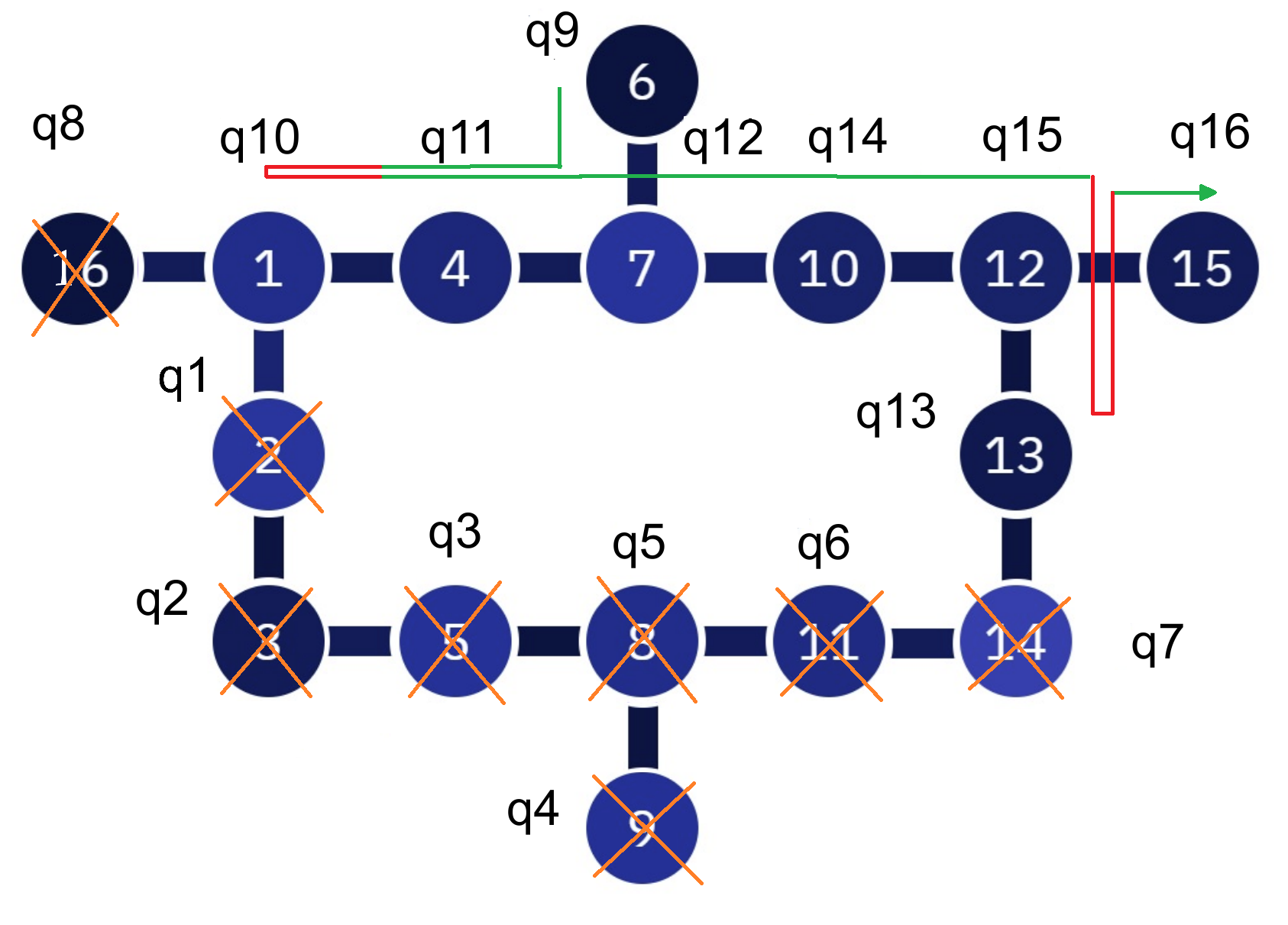}
\caption{
The graph for the 16-qubit ``sun'' architecture. The $9$-th cascade that excludes the vertex that corresponds to the logical qubits with indexes from the $\{1,\dots,8\}$ set.}
\end{figure}

\begin{figure}[H]
\includegraphics[width=0.28\textwidth]{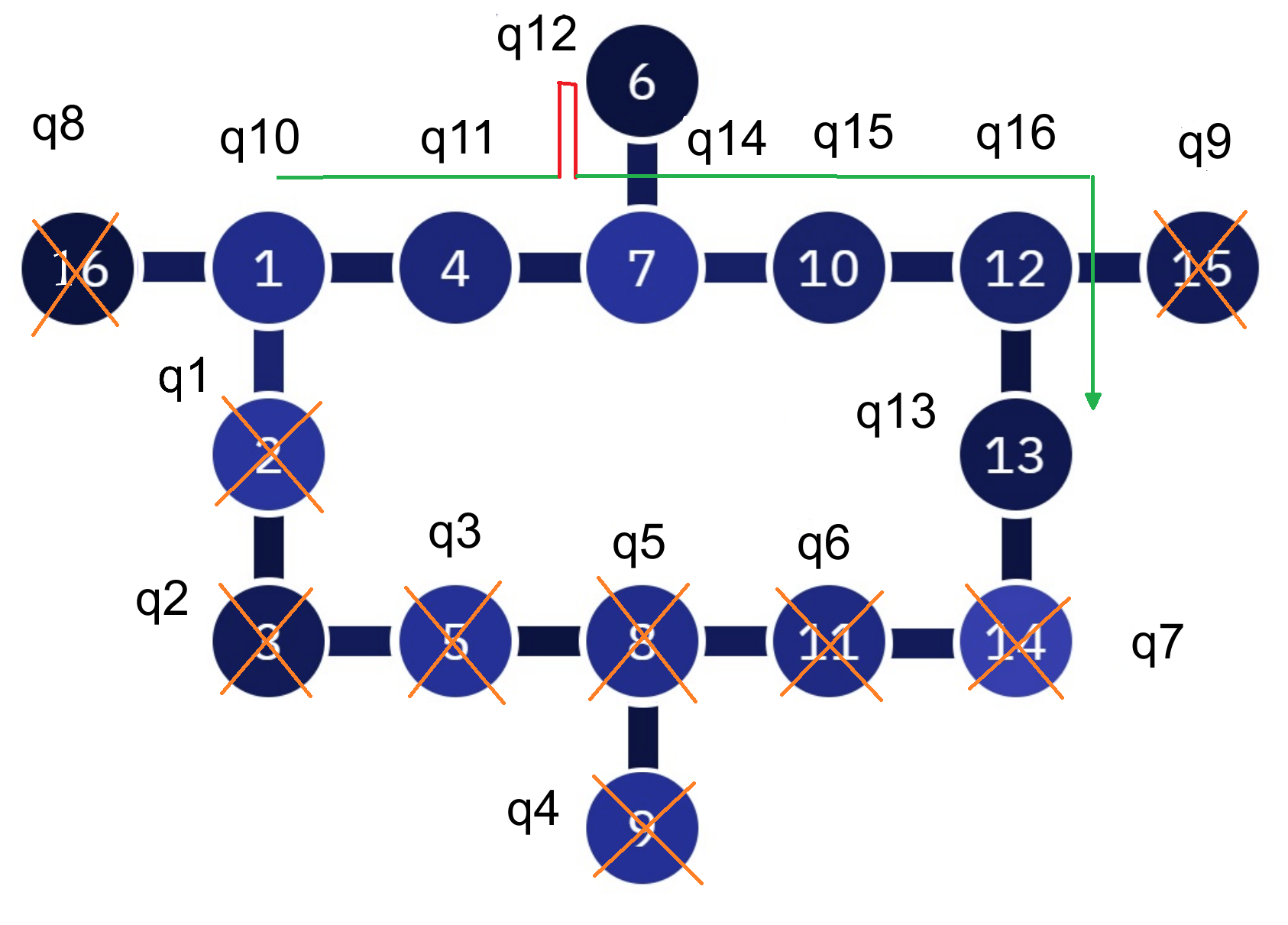}
\caption{
The graph for the 16-qubit ``sun'' architecture. The $10$-th cascade that excludes the vertex that corresponds to the logical qubits with indexes from the $\{1,\dots,9\}$ set.}
\end{figure}

\begin{figure}[H]
\includegraphics[width=0.28\textwidth]{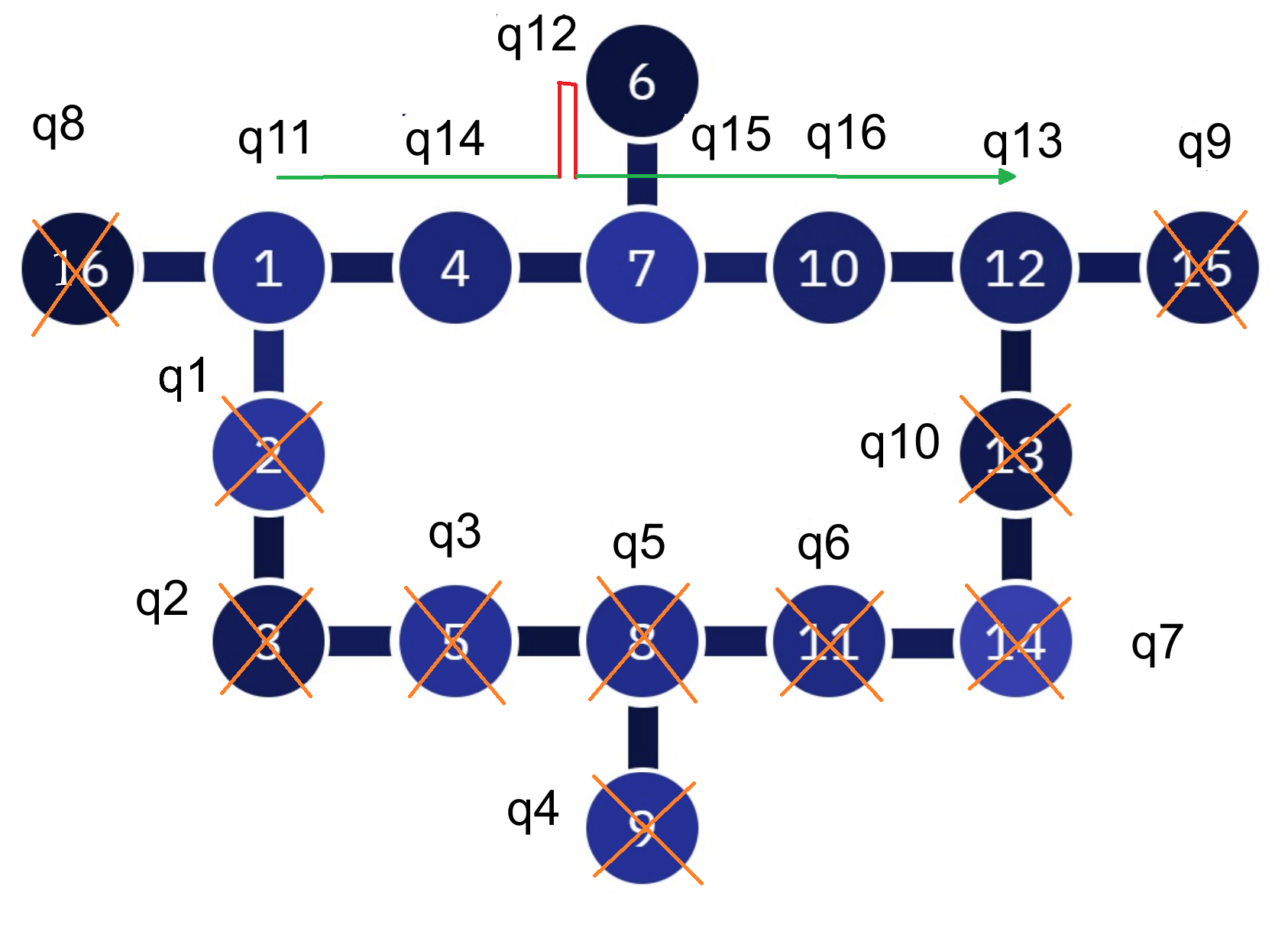}
\caption{
The graph for the 16-qubit ``sun'' architecture. The $11$-th cascade excludes the vertex that corresponds to the logical qubits with indexes from the $\{1,\dots,10\}$ set.}
\end{figure}

\begin{figure}[H]
\includegraphics[width=0.28\textwidth]{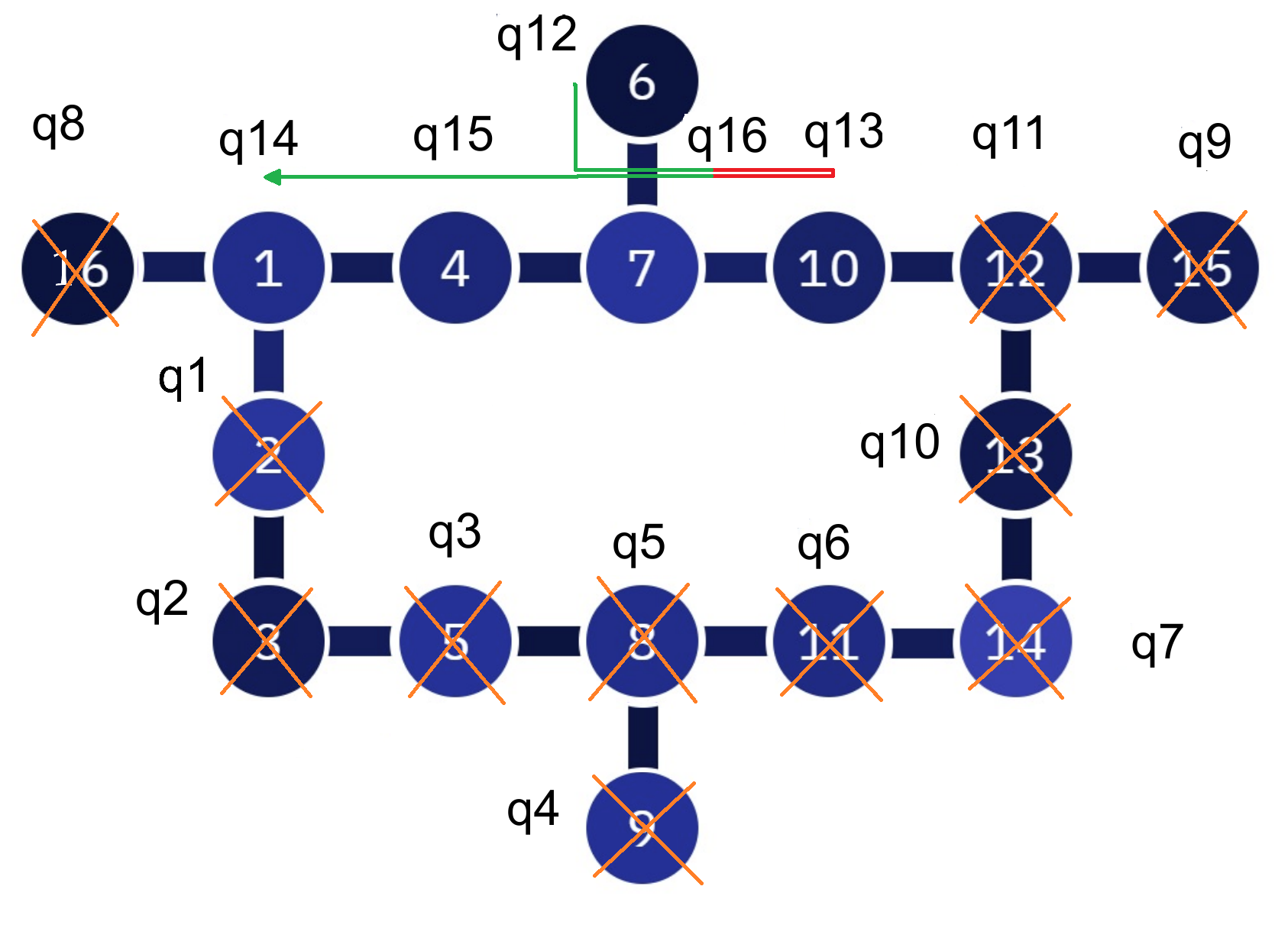}
\caption{
The graph for the 16-qubit ``sun'' architecture. The $12$-th cascade excludes the vertex that corresponds to the logical qubits with indexes from the $\{1,\dots,11\}$ set.}
\end{figure}

\begin{figure}[H]
\includegraphics[width=0.28\textwidth]{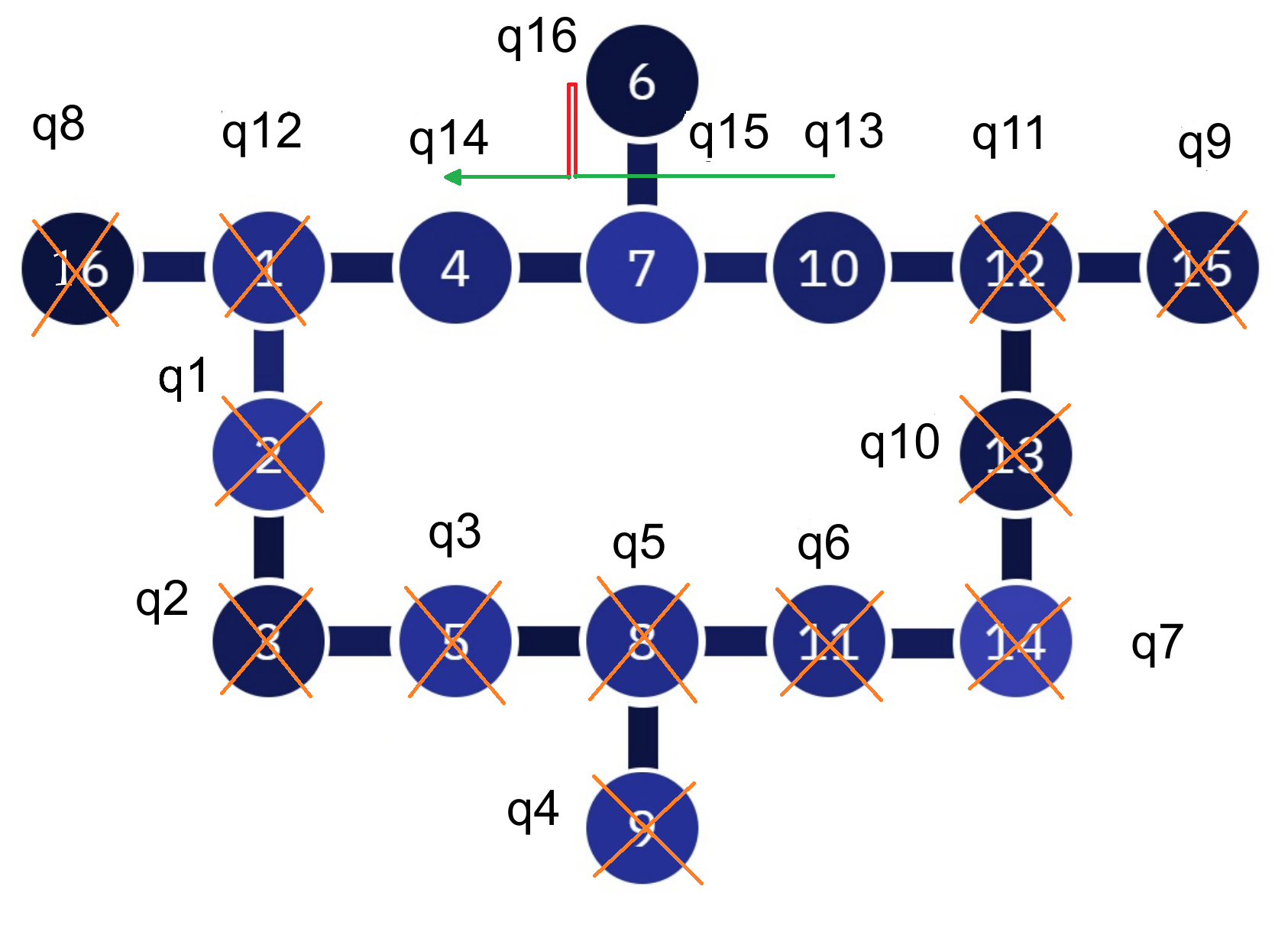}
\caption{
The graph for the 16-qubit ``sun'' architecture. The $13$-th cascade that excludes the vertex that corresponds to the logical qubits with indexes from the $\{1,\dots,12\}$ set.}
\end{figure}

\begin{figure}[H]
\includegraphics[width=0.28\textwidth]{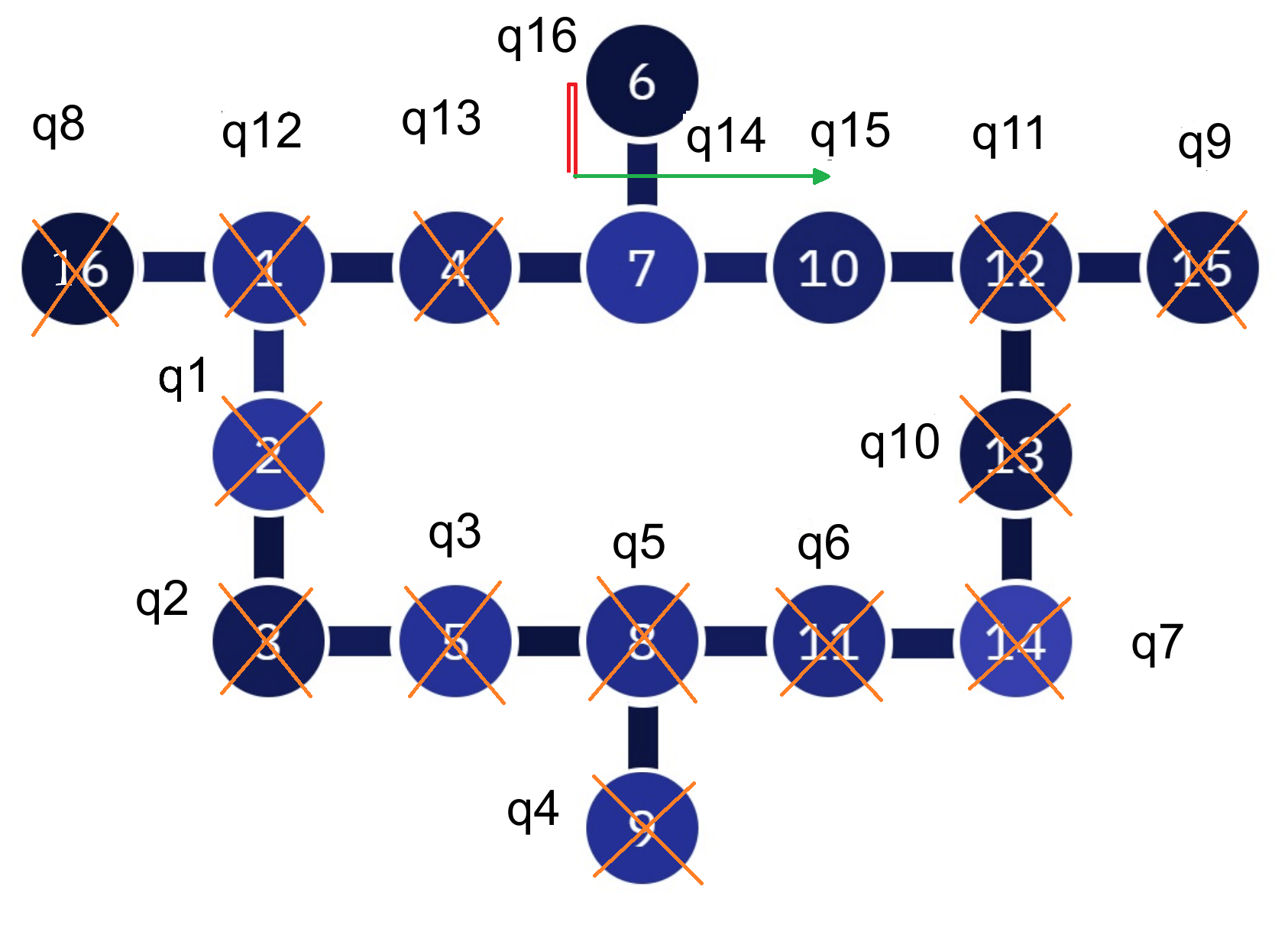}
\caption{
The graph for the 16-qubit ``sun'' architecture. The $14$-th cascade excludes the vertex that corresponds to the logical qubits with indexes from the $\{1,\dots,13\}$ set.}
\end{figure}

\begin{figure}[H]
\includegraphics[width=0.28\textwidth]{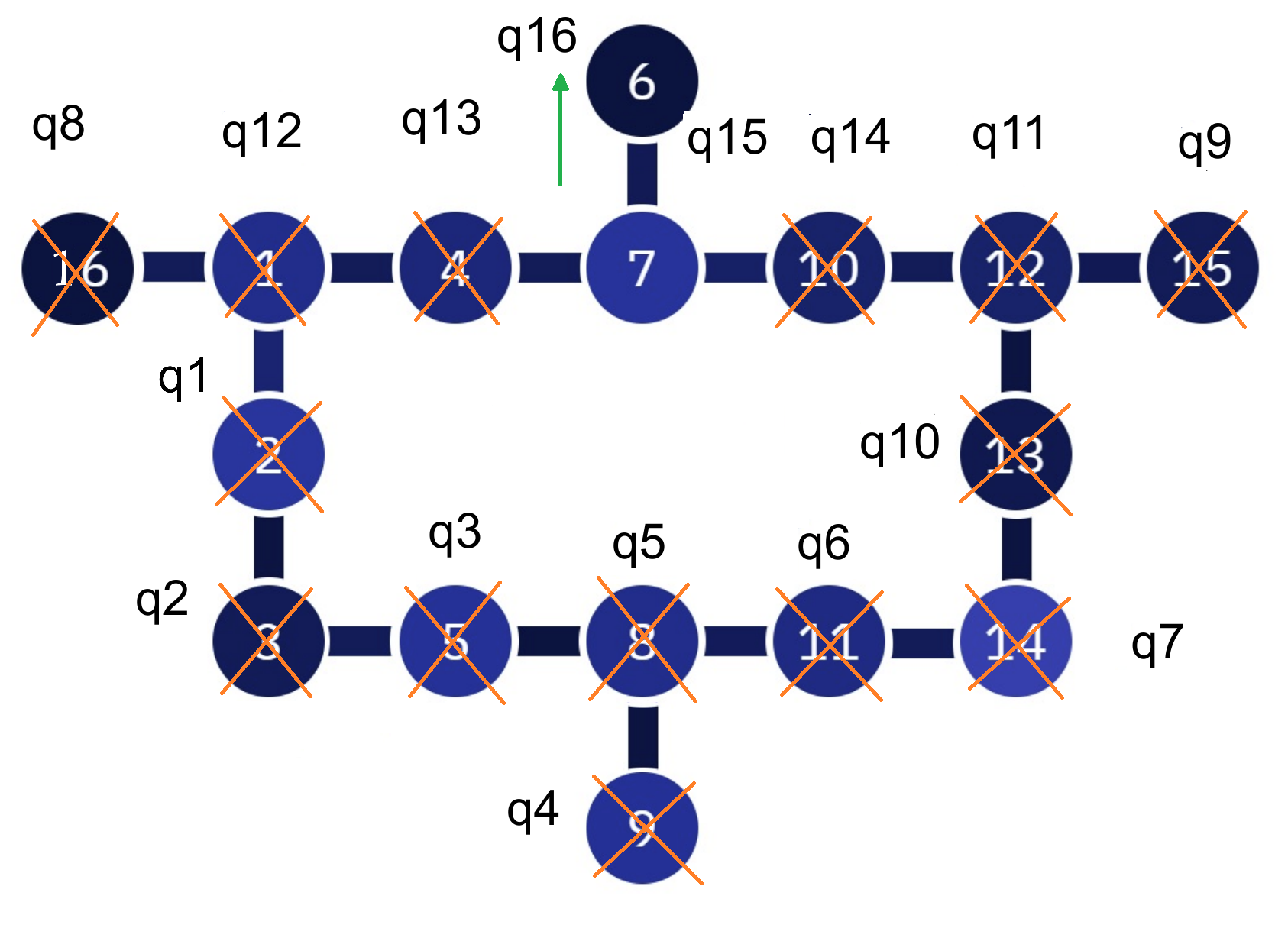}
\caption{
The graph for the 16-qubit ``sun'' architecture. The $15$-th cascade excludes the vertex that corresponds to the logical qubits with indexes from the $\{1,\dots,14\}$ set.}
\end{figure}

\begin{figure}[H]
\includegraphics[width=0.28\textwidth]{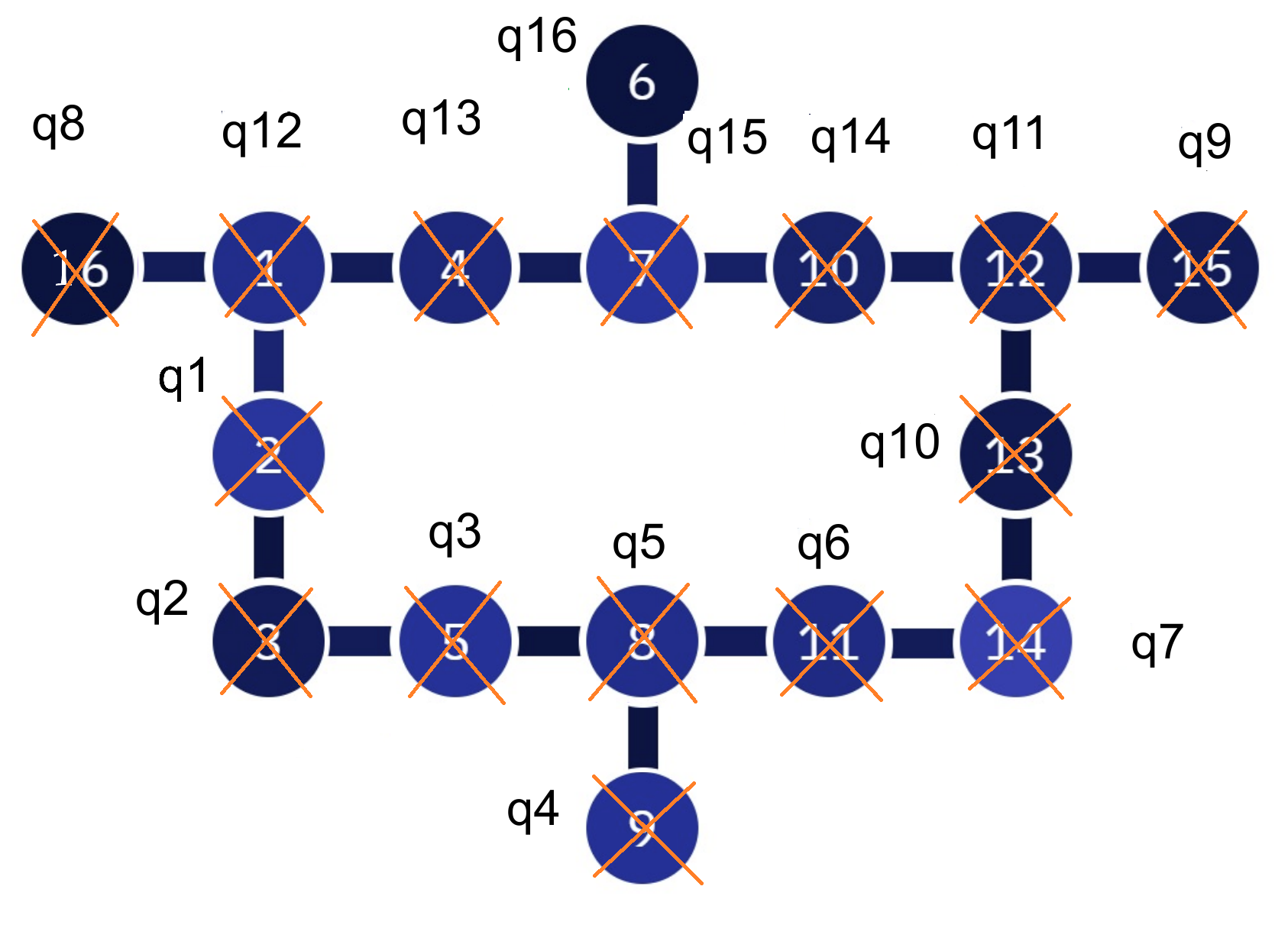}
\caption{
The graph for the 16-qubit ``sun'' architecture. The $16$-th cascade that excludes the vertex that corresponds to the logical qubits with indexes from the $\{1,\dots,15\}$ set.}
\end{figure}

\bibliography{tcs}

\end{document}